\documentclass[a4paper,12pt]{article}
\usepackage{expl3}
\usepackage{amsmath,amssymb,mathrsfs,amsthm,tikz,shuffle,paralist}
\usepackage{dsfont}
\usepackage{amsbsy}

\usepackage{xcolor}
\usepackage{relsize}
\usepackage{subfig}
\usepackage{comment}

\usepackage[font={small},labelfont=bf]{caption}

\usepackage[compat=1.1.0]{tikz-feynman}
\usetikzlibrary{shapes.misc}
\usetikzlibrary{cd}
\usepackage[size=scriptsize]{todonotes}
\usepackage{stackengine}
\usepackage[bottom]{footmisc}
\usepackage{booktabs} 

\usetikzlibrary{arrows}

\newcommand{\LS}{{\text{LS}}}
\newcommand{\diff}{{d}}

\newcommand{\sectora}{a}
\newcommand{\sectorb}{b}
\newcommand{\sectorc}{c}
\newcommand{\sectord}{d}

\newcommand{\bq}{\begin{eqnarray}}
\newcommand{\eq}{\end{eqnarray}}
\newcommand{\eps}{\varepsilon}
\newcommand{\NF}{n}

\allowdisplaybreaks

\usepackage{url}

\usepackage{breakurl}
\usepackage[colorlinks=true
,urlcolor=blue
,anchorcolor=blue
,citecolor=blue
,filecolor=blue
,linkcolor=blue
,menucolor=blue
,pagecolor=blue
,linktocpage=true
,pdfproducer=medialab
,pdfa=true
]{hyperref}
\hypersetup{breaklinks=true}

\usepackage{tcolorbox}
\usepackage{nicematrix}

\newcommand{\integralFinalReally}{\tilde{\mathcal{J}}}
\newcommand{\integralFinal}{\mathcal{J}}
\newcommand{\integralprec}{\mathcal{I}}
\newcommand{\integralseed}{\mathcal{I}_{\text{seed}}}

\usepackage[utf8]{inputenc}
\usepackage{graphicx}

\usepackage{jheppub}

\makeatletter
\DeclareRobustCommand*{\bfseries}{%
  \not@math@alphabet\bfseries\mathbf
  \fontseries\bfdefault\selectfont
  \boldmath
}

\begin{document}

\title{Calabi--Yau Feynman integrals in gravity: \\ $\varepsilon$-factorized form for apparent singularities}

\preprint{MITP-24-089\ PSI-PR-25-01}

\author[a]{Hjalte Frellesvig,}
\emailAdd{hjalte.frellesvig@nbi.ku.dk}
\author[a]{Roger Morales,}
\emailAdd{roger.morales@nbi.ku.dk}
\author[b]{Sebastian P\"ogel,}
\emailAdd{sebastian.poegel@psi.ch}
\author[c]{Stefan Weinzierl,}
\emailAdd{weinzierl@uni-mainz.de}
\author[a,d]{and Matthias Wilhelm}
\emailAdd{mwilhelm@imada.sdu.dk}

\affiliation[a]{%
Niels Bohr International Academy, Niels Bohr Institute, Copenhagen University, Blegdamsvej 17, DK-2100 Copenhagen \O{}, Denmark}
\affiliation[b]{%
Paul Scherrer Institut, CH-5232 Villigen PSI, Switzerland}
\affiliation[c]{%
PRISMA Cluster of Excellence, Institut für Physik, Johannes Gutenberg-Universität Mainz, D-55099 Mainz, Germany}
\affiliation[d]{%
Center for Quantum Mathematics, Department of Mathematics and Computer Science, University of Southern Denmark, DK-5230 Odense M, Denmark}

\date{\today}%

\abstract{%
We study a recently identified four-loop Feynman integral that contains a three-dimensional Calabi--Yau geometry and contributes to the scattering of black holes in classical gravity at fifth post-Minkowskian and second self-force order (5PM 2SF) in the conservative sector. In contrast to previously studied Calabi--Yau Feynman integrals, the higher-order differential equation that this integral satisfies in dimensional regularization exhibits $\varepsilon$-dependent apparent singularities. We introduce an appropriate ansatz which allows us to bring such cases into an $\varepsilon$-factorized form.
As a proof of principle, we apply it to the integral at hand.}

\maketitle

\section{Introduction} 
\label{sec:intro}

Feynman integrals are key ingredients for the calculation of high-precision predictions using perturbation theory, with applications ranging from particle physics to gravitational waves emitted during the inspiral of black-hole binary systems. 

A crucial feature of Feynman integrals are the intricate geometries they contain, which completely determine the space of functions to which the integrals evaluate; see ref.~\cite{Bourjaily:2022bwx} for a recent review. The simplest Feynman integrals can be written in terms of so-called multiple polylogarithms~\cite{Chen:1977oja,Goncharov:1995ifj}, which are iterated integrals over the Riemann sphere generalizing classical polylogarithms. However, more recently, examples of Feynman integrals have been identified which depend on elliptic curves~\cite{Sabry:1962rge,Broadhurst:1993mw,Laporta:2004rb,Caron-Huot:2012awx,Adams:2013nia,Bloch:2013tra,Adams:2014vja,Remiddi:2016gno,Adams:2016xah,Broedel:2017siw,Kristensson:2021ani,Giroux:2022wav,Morales:2022csr,McLeod:2023qdf,Stawinski:2023qtw,Giroux:2024yxu,Spiering:2024sea} and higher-genus curves~\cite{Huang:2013kh,Georgoudis:2015hca,Marzucca:2023gto} as well as on Calabi--Yau  varieties\footnote{Calabi--Yau varieties can always be desingularized to obtain Calabi--Yau manifolds; however, this does not play a role in the context of Feynman integrals.}~\cite{Brown:2010bw,Bourjaily:2018ycu,Bourjaily:2018yfy,Bonisch:2021yfw,Broedel:2021zij,Lairez:2022zkj,Pogel:2022vat,Duhr:2022dxb,Cao:2023tpx,McLeod:2023doa,Duhr:2023eld,Duhr:2024hjf}, for which the corresponding spaces of functions are much less understood. Remarkably, Feynman integrals in classical gravity have also been found to contain K3 surfaces giving rise to elliptic integrals at three loops~\cite{Bern:2021dqo,Dlapa:2021npj,Dlapa:2022wdu}, as well as three-dimensional Calabi--Yau varieties at four loops~\cite{Frellesvig:2023bbf,Klemm:2024wtd,Driesse:2024feo}. These non-trivial geometries thus pose a fundamental challenge for calculating Feynman integrals and obtaining high-precision theoretical predictions.

Among the most prominent methods for calculating Feynman integrals are differential equations~\cite{Kotikov:1990kg,Kotikov:1991hm,Kotikov:1991pm,Gehrmann:1999as}. 
Via integration-by-parts (IBP) identities~\cite{Chetyrkin:1981qh}, all Feynman integrals are known to be reducible to a finite number of master integrals~\cite{Smirnov:2010hn,Bitoun:2017nre}, which can be organized into sectors of integrals sharing the same set of propagators. 
The master integrals satisfy a linear system of first-order differential equations.
This system of first-order differential equations naturally has a block-triangular structure, 
induced by the sectors.
By taking further derivatives, we may obtain for any (master) integral a single higher-order differential equation, where the corresponding higher-order differential operator is the so-called Picard--Fuchs operator of the integral. 
The essential information is already captured by considering the Picard--Fuchs operator on the maximal cut, i.e.\ ignoring subsectors.
Throughout this paper, we adopt the convention that by the Picard--Fuchs operator of a Feynman integral
we understand the Picard--Fuchs operator of the integral on the maximal cut.
For multiple polylogarithms, the Picard--Fuchs operator factorizes into a product of operators of order one, while it yields a higher-order irreducible operator for more complicated geometries. This way, the Picard--Fuchs operator can be used to identify and characterize the geometry that the integrals depend on.

Within the method of differential equations, and under dimensional regularization, a Feynman integral can be considered to be solved once it is brought into $\varepsilon$-factorized or \emph{canonical} form, in which case the integral can be systematically computed order-by-order in the dimensional regulator $\varepsilon$. The canonical form was first introduced in ref.~\cite{Henn:2013pwa} for Feynman integrals evaluating to multiple polylogarithms. In the polylogarithmic case, systematic algorithms exist which can bring the differential equation into canonical form by performing a change of basis \cite{Lee:2014ioa}, implemented in a number of computer programs \cite{Prausa:2017ltv,Gituliar:2017vzm,Meyer:2017joq,Dlapa:2020cwj,Lee:2020zfb}. More recently, the notion of a canonical form has been extended to Feynman integrals involving elliptic curves~\cite{Adams:2018yfj,Dlapa:2020cwj,Dlapa:2022wdu,Jiang:2023jmk,Gorges:2023zgv}, higher-genus curves \cite{Duhr:2024uid} and the simplest cases of Calabi--Yau varieties~\cite{Pogel:2022ken,Pogel:2022vat}, building on the knowledge of the underlying geometry via the Picard--Fuchs operator. However, the latter approaches for Calabi--Yau integrals require a good choice of master integrals and are limited in what classes of Picard--Fuchs operators they can be applied to, as we will show.

In this paper, we generalize the method of ref.~\cite{Pogel:2022vat} to bring a considerably larger class of Feynman integral differential equations into canonical form. Namely, we also allow for Picard--Fuchs operators containing $\varepsilon$-dependent apparent singularities, for which it is necessary to enlarge the ansatz for the transformation matrix to canonical form. An apparent singularity is a singularity of an operator at which all solutions are non-singular, thus it has no physical origin. Typically, singularities at $\varepsilon$-dependent values of the kinematics are apparent since nothing physical is expected to happen there. 
The fact that the regulator $\varepsilon$ only affects such unphysical singularities was already observed in ref.~\cite{delaCruz:2024xit}.
In general, $\varepsilon$-dependent apparent singularities can occur both in the linear system of first-order differential equations and in the Picard--Fuchs operator. In the case of linear systems, $\varepsilon$-dependent apparent singularities can be algorithmically removed by performing a transformation on the master integral basis, see e.g.\ refs.~\cite{10.1145/220346.220385,10.1145/2755996.2756668} for computer program implementations. By contrast, $\varepsilon$-dependent apparent singularities can appear in the Picard--Fuchs operator and at the same time be absent in the linear system. While there exist methods to remove $\varepsilon$-dependent apparent singularities in Picard--Fuchs operators~\cite{TSAI2000747,Abramov:2004,CHEN2016617,Slavyanov:2016,10.1016/j.jsc.2019.02.009}, with computer implementations in \texttt{Maple}~\cite{Abramov:2004} and \texttt{SageMath}~\cite{Kauers:2013juk}, they are not systematic or generate a differential operator of higher order than the original. In many cases, the presence of an $\varepsilon$-dependent apparent singularity in the Picard--Fuchs operator signals a suboptimal choice of master integrals, or the requirement of choosing more than one suitable starting integral to construct a complete set of master integrals using a derivative basis. However, there is no systematic procedure to find the optimal basis of master integrals, and no guarantee that a choice of master integrals free from $\varepsilon$-dependent apparent singularities can be found. With our extended method that accommodates for operators with $\varepsilon$-dependent apparent singularities, we can therefore relax the necessity of finding a good choice of master integrals to derive a canonical form.
	In cases where typically more than one starting integral is required we will see that such additional integrals are automatically included.

\begin{figure}[tb]
\centering
\subfloat[]{\begin{tikzpicture}[baseline={([yshift=-0
cm]current bounding box.center)}, scale=1.4] 
	\node[] (a) at (0,0) {};
	\node[] (a1) at (0.5,0) {};
	\node[] (a2) at (1,0) {};
	\node[] (b) at (0,-1) {};
	\node[] (b1) at (0.5,-1) {};
	\node[] (b2) at (1,-1) {};
	\node[] (p1) at ($(a)+(-0.2,0)$) {};
	\node[] (p2) at ($(b)+(-0.2,0)$) {};
	\node[] (p3) at ($(b2)+(0.2,0)$) {};
	\node[] (p4) at ($(a2)+(0.2,0)$) {};
	\draw[line width=0.15mm] (b.center) -- (a2.center);
	\draw[line width=0.15mm] (b1.center) -- (a1.center);
	\draw[line width=0.15mm] (b2.center) -- (a.center);
	\draw[line width=0.5mm] (p1.center) -- (p4.center);
	\draw[line width=0.5mm] (p2.center) -- (p3.center);
\end{tikzpicture}} \quad
\subfloat[]{\begin{tikzpicture}[baseline={([yshift=-0
cm]current bounding box.center)}, scale=1.4] 
	\node[] (a) at (0,0) {};
	\node[] (a1) at (0.5,0) {};
	\node[] (a2) at (1,0) {};
	\node[] (c) at (0,-1) {};
	\node[] (c1) at (0.5,-1) {};
	\node[] (c2) at (1,-1) {};
	\node[] (p1) at ($(a)+(-0.2,0)$) {};
	\node[] (p2) at ($(c)+(-0.2,0)$) {};
	\node[] (p3) at ($(c2)+(0.2,0)$) {};
	\node[] (p4) at ($(a2)+(0.2,0)$) {};
	\draw[line width=0.15mm] (c.center) -- (a.center);
	\draw[line width=0.15mm] (c.center) -- (a1.center);
	\draw[line width=0.15mm] (c2.center) -- (a2.center);
	\draw[line width=0.15mm] (c.center) -- (a2.center);
	\draw[line width=0.15mm] (c1.center) -- (a2.center);
	\draw[line width=0.5mm] (p1.center) -- (p4.center);
	\draw[line width=0.5mm] (p2.center) -- (p3.center);
\end{tikzpicture}}\quad
\subfloat[]{\begin{tikzpicture}[baseline={([yshift=-0
cm]current bounding box.center)}, scale=1.4] 
	\node[] (a) at (0,0) {};
	\node[] (a1) at (0.5,0) {};
	\node[] (a2) at (1,0) {};
	\node[] (b) at (1,-0.5) {};
	\node[] (c) at (0,-1) {};
	\node[] (c1) at (0.5,-1) {};
	\node[] (c2) at (1,-1) {};
	\node[] (p1) at ($(a)+(-0.2,0)$) {};
	\node[] (p2) at ($(c)+(-0.2,0)$) {};
	\node[] (p3) at ($(c2)+(0.2,0)$) {};
	\node[] (p4) at ($(a2)+(0.2,0)$) {};
	\draw[line width=0.15mm] (c.center) -- (a.center);
	\draw[line width=0.15mm] (c.center) -- (a1.center);
	\draw[line width=0.15mm] (c2.center) -- (a2.center);
	\draw[line width=0.15mm] (c.center) -- (b.center);
	\draw[line width=0.15mm] (c1.center) -- (b.center);
	\draw[line width=0.5mm] (p1.center) -- (p4.center);
	\draw[line width=0.5mm] (p2.center) -- (p3.center);
\end{tikzpicture}} 
\quad
\subfloat[]{\begin{tikzpicture}[baseline={([yshift=-0
cm]current bounding box.center)}, scale=1.4] 
	\node[] (a) at (0,0) {};
	\node[] (a1) at (0.5,0) {};
	\node[] (a2) at (1,0) {};
	\node[] (b) at (0,-0.5) {};
	\node[] (b1) at (1,-0.5) {};
	\node[] (c) at (0,-1) {};
	\node[] (c1) at (0.5,-1) {};
	\node[] (c2) at (1,-1) {};
	\node[] (p1) at ($(a)+(-0.2,0)$) {};
	\node[] (p2) at ($(c)+(-0.2,0)$) {};
	\node[] (p3) at ($(c2)+(0.2,0)$) {};
	\node[] (p4) at ($(a2)+(0.2,0)$) {};
	\draw[line width=0.15mm] (c.center) -- (a.center);
	\draw[line width=0.15mm] (b.center) -- (b1.center);
	\draw[line width=0.15mm] (c2.center) -- (a2.center);
	\draw[line width=0.15mm] (b1.center) -- (c1.center);
	\draw[line width=0.15mm] (b.center) -- (a1.center);
	\draw[line width=0.5mm] (p1.center) -- (p4.center);
	\draw[line width=0.5mm] (p2.center) -- (p3.center);
\end{tikzpicture}}
\caption{The four-loop Feynman integral containing the Calabi--Yau three-variety (d) and its classical subsectors (a)--(c). The thin lines denote quadratic (graviton) propagators while the thick lines denote linearized massive (scalar) propagators.}
\label{fig: diagrams}
\end{figure}
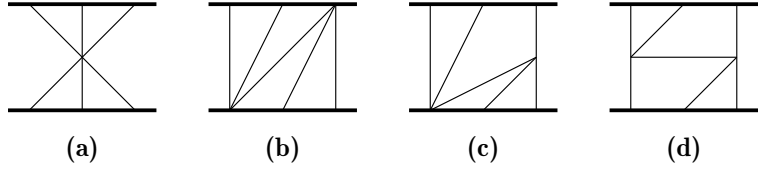

Our prime example for a Picard--Fuchs operator with an $\varepsilon$-dependent apparent singularity is a particular four-loop single-scale Feynman integral that contains a Calabi--Yau three-variety; see fig.~\ref{fig: diagrams}(d). This Calabi--Yau integral, which was recently identified~\cite{Frellesvig:2023bbf} by analyzing its leading singularity, contributes to the classical dynamics and gravitational waves emitted during the coalescence of binary systems of black holes and neutron stars in the post-Minkowskian (PM) expansion at 5PM order; see refs.~\cite{Damour:2016gwp,Bern:2019crd,Kalin:2020mvi,Mogull:2020sak} for more details as well as refs.~\cite{Bjerrum-Bohr:2022blt,Buonanno:2022pgc} for a review. Specifically, it appears in the conservative sector at second order in the self-force expansion (2SF). This lies beyond the current state of the art in the literature at 5PM, which is the 1SF order~\cite{Driesse:2024xad,Driesse:2024feo}, and thus it belongs to the next sector that needs to be calculated in PM theory. With the purpose of facilitating the calculation of the 5PM 2SF conservative sector in classical gravity, as well as of further Calabi--Yau integrals that will appear in the PM expansion in the future, in this paper we showcase how to bring the differential equation for this integral into canonical form.

While the prime example in our discussion will be a Calabi--Yau Feynman integral, the procedure we describe is not dependent on the underlying geometry.
In fact, it relies solely on knowledge of the Picard--Fuchs operator for a chosen seed integral, and is entirely agnostic of the associated geometry.
Thus we suspect that our procedure generalizes to geometries beyond those of Calabi--Yau type.

The remainder of this paper is structured as follows. After motivating the study of $\varepsilon$-dependent apparent singularities in section \ref{sec:motivation}, we describe in section \ref{sec:analysis_CY} the aforementioned Feynman integral in classical gravity. We find that its associated Picard--Fuchs operator indeed has an $\varepsilon$-dependent apparent singularity, and we further characterize the geometry on which it depends. We then discuss in section \ref{sec:eps-form_general} how to bring a differential equation into canonical form in cases where the corresponding Picard--Fuchs operator contains $\varepsilon$-dependent apparent singularities, detailing the general method and illustrating it with a simple example. In section \ref{sec:eps-form_CY}, we apply this procedure to the gravity integral containing the $\varepsilon$-dependent apparent singularity, both at the level of the topsector as well as including its classical subsectors; see fig.~\ref{fig: diagrams}. We conclude with a summary and outlook on future directions in section~\ref{sec:conclusions}.

\section{Motivation} 
\label{sec:motivation}

In this section, we introduce $\eps$-dependent apparent singularities, which are the main subject of the present paper. 

We start with a toy example for Feynman integrals depending on a single kinematic variable $x$. Let us define $\mathcal{J}=(\mathcal{J}_1,\mathcal{J}_2,\mathcal{J}_3)^T$ as a vector of three master integrals that satisfy a linear system of $\varepsilon$-factorized differential equations. In particular, we consider the following linear system:
\bq
\label{toy_example_linear_system_3}
 \frac{d}{dx} 
 \left( \begin{array}{c}
  {\mathcal J}_1 \\
  {\mathcal J}_2 \\
  {\mathcal J}_3 \\
 \end{array} \right)
 & = &
 \eps \left[
  \left( \begin{array}{rrr}
   -9 & -33 & 56 \\
   \frac{9}{2} & 18 & -33 \\
   1 & \frac{9}{2} & -9 \\ 
  \end{array} \right)
  \frac{1}{x}
  +
  \left( \begin{array}{rrr}
   12 & 40 & -64 \\
   -7 & -24 & 40 \\
   -2 & -7 & 12 \\
  \end{array} \right)
  \frac{1}{x-1} 
  \right]
 \left( \begin{array}{c}
  {\mathcal J}_1 \\
  {\mathcal J}_2 \\
  {\mathcal J}_3 \\
 \end{array} \right).
\eq
The finite singular points of this system of differential equations are clearly at $x=0$ and $x=1$.
It is easily checked that for generic $x$ the vector space generated by $\{{\mathcal J}_1,{\mathcal J}_2,{\mathcal J}_3\}$ is also spanned by
\bq
\label{example_derivative_basis}
 \left\{ {\mathcal J}_1, \frac{d}{dx} {\mathcal J}_1, \frac{d^2}{dx^2}{\mathcal J}_1 \right\}.
\eq
This implies that ${\mathcal J}_1$ satisfies a third-order linear differential equation
\bq
\label{toy_example_Picard_Fuchs_3}
\mathcal{L}_3 \, {\mathcal J}_1 \equiv \left[
 C_3\left(x,\eps\right) \frac{d^3}{dx^3}
 + C_2\left(x,\eps\right) \frac{d^2}{dx^2}
 + C_1\left(x,\eps\right) \frac{d}{dx}
 + C_0\left(x,\eps\right) 
 \right] {\mathcal J}_1
 & = & 0\,,
\eq
where $\mathcal{L}_3$ is its Picard--Fuchs operator, with
\bq
 C_3\left(x,\eps\right)
 & = &
 x^2 \left(x-1\right)^2 \left[ 128 x \left(1-x\right) + \left(x-9\right) \left(x^2+46x+17\right) \eps \right].
\eq
The remaining coefficients $C_2(x,\eps)$--$C_0(x,\eps)$ are similarly polynomials in $x$ and $\eps$,  but their explicit expressions will not be relevant in the following.
The singularities of the third-order differential equation~(\ref{toy_example_Picard_Fuchs_3})
are given by the leading coefficient $C_3(x,\eps)$.
We recognise the singular points $x=0$ and $x=1$ already encountered in the linear system of eq.~(\ref{toy_example_linear_system_3}),
but in addition there are singular points determined as the three solutions to
\bq
\label{example_eps_dependent_apparent_singularity}
 128 x \left(1-x\right) + \left(x-9\right) \left(x^2+46x+17\right) \eps
 & = & 0\,.
\eq
These are $\eps$-dependent apparent singularities.
As the linear system of eq.~(\ref{toy_example_linear_system_3}) is regular at these points, the solutions of the linear system
will be regular as well, hence these singularities are apparent singularities.
The locus of these singularities in $x$-space depends on the parameter $\eps$, hence they are $\eps$-dependent apparent singularities.
These singularities are the focus of this paper.
They enter through the transformation from the basis $({\mathcal J}_1,{\mathcal J}_2,{\mathcal J}_2)^T$ to the basis given in eq.~(\ref{example_derivative_basis}). For generic $x$ this is an invertible transformation, but at the apparent singularities the three master integrals of the derivative basis in
eq.~(\ref{example_derivative_basis}) become linearly dependent.
This is easily seen from the determinant of the transformation matrix, which is given by
\bq
 \frac{\eps^2 \left[ 128 x (1-x) + (x-9) (x^2+46x+17) \eps \right]}{x^3 (1-x)^3}\,,
\eq
and vanishes at the apparent singularities.
We can repeat the above analysis with ${\mathcal J}_2$ or ${\mathcal J}_3$ instead of ${\mathcal J}_1$, arriving at the same conclusion:
The Picard--Fuchs operator for any of the three master integrals ${\mathcal J}_1$, ${\mathcal J}_2$ or ${\mathcal J}_3$
has $\eps$-dependent apparent singularities.

The above toy example is simpler than the integrals we address in this paper: 
In this toy example we can choose a different basis, where neither the linear system nor the Picard--Fuchs operators for all
three master integrals have $\eps$-dependent apparent singularities.
To see this, consider the new basis
\bq
 \left( \begin{array}{c}
  {\mathcal J}_1' \\
  {\mathcal J}_2' \\
  {\mathcal J}_3' \\
 \end{array} \right)
 & = &
  \left( \begin{array}{rrr}
   -\frac{1}{2} & -1 & 1 \\
   1 & 3 & -4 \\
   1 & 4 & -8 \\ 
  \end{array} \right)
 \left( \begin{array}{c}
  {\mathcal J}_1 \\
  {\mathcal J}_2 \\
  {\mathcal J}_3 \\
 \end{array} \right).
\eq
In this new basis the linear system reads
\bq
 \frac{d}{dx} 
 \left( \begin{array}{c}
  {\mathcal J}_1' \\
  {\mathcal J}_2' \\
  {\mathcal J}_3' \\
 \end{array} \right)
 & = &
 \eps \left[
  \left( \begin{array}{rrr}
   0 & 1 & 0 \\
   1 & 0 & 1 \\
   0 & 1 & 0 \\ 
  \end{array} \right)
  \frac{1}{x}
  +
  \left( \begin{array}{rrr}
   0 & -1 & 0 \\
   0 & 0 & -1 \\
   0 & 0 & 0 \\
  \end{array} \right)
  \frac{1}{x-1} 
  \right]
 \left( \begin{array}{c}
  {\mathcal J}_1' \\
  {\mathcal J}_2' \\
  {\mathcal J}_3' \\
 \end{array} \right),
\eq
and it is easily checked that the only (finite) singularities of the Picard--Fuchs operators are at $x=0$ and $x=1$ for all ${\mathcal J}_1'$, ${\mathcal J}_2'$
and ${\mathcal J}_3'$.

Let us consider a second toy example, given by the following system of differential equations
\bq
\label{toy_example_linear_system_4}
 \frac{d}{dx} 
 \left( \begin{array}{c}
  {\mathcal J}_1 \\
  {\mathcal J}_2 \\
  {\mathcal J}_3 \\
 \end{array} \right)
 & = &
 \eps 
  \left( \begin{array}{ccc}
   \frac{1}{x-2} & \frac{1}{x-1} & \frac{1}{x} \\
   \frac{1}{x-4} & \frac{1}{x-3} & \frac{1}{x-1} \\
   \frac{1}{x-5} & \frac{1}{x-4} & \frac{1}{x-2} \\
  \end{array} \right)
 \left( \begin{array}{c}
  {\mathcal J}_1 \\
  {\mathcal J}_2 \\
  {\mathcal J}_3 \\
 \end{array} \right).
\eq
The finite singular points of this system of differential equations are at $x\in\{0,\dots,5\}$.
If we look at the Picard--Fuchs operators for the integrals ${\mathcal J}_1$, ${\mathcal J}_2$ or ${\mathcal J}_3$,
we find again $\eps$-dependent apparent singularities.
In contrast to the first example, we can show for this case that there is no 
constant $\mathrm{GL}(3,{\mathbb C})$-rotation\footnote{The solutions in the basis ${\mathcal J}$ are of uniform weight (for boundary constants of uniform weight)
and a constant $\mathrm{GL}(3,{\mathbb C})$-rotation preserves this property.}
to a basis $\{ {\mathcal J}_1', {\mathcal J}_2', {\mathcal J}_3' \}$ such that the Picard--Fuchs operator for ${\mathcal J}_1'$ is
free of $\eps$-dependent apparent singularities.
(This implies that the Picard--Fuchs operators of 
all three integrals ${\mathcal J}_1'$, ${\mathcal J}_2'$ and ${\mathcal J}_3'$ will have
$\eps$-dependent apparent singularities.)

From the two examples above, we learn two lessons:
\begin{enumerate}
\item An $\eps$-dependent apparent singularity in the Picard--Fuchs operator for a master integral ${\mathcal J}_i$ 
does not exclude this master integral from appearing in a basis for an $\eps$-factorized differential equation.  
\item It may or may not be the case that there exists a basis such that at least one master integral has a Picard--Fuchs operator
without $\eps$-dependent apparent singularities.
\end{enumerate}
Our goal is not to decide if there exists a basis such that at least one master integral has a Picard--Fuchs operator
without $\eps$-dependent apparent singularities.
Our main goal is to construct an $\eps$-factorized basis ${\mathcal J}$ from a pre-canonical basis ${\mathcal I}$.
We usually work with a seed integral $\integralseed$ and normalize it by an $\eps$-independent function $\omega(x)$ related to the leading singularity of $\integralseed$ to obtain a candidate for one integral from the $\eps$-factorized basis ${\mathcal J}$,
\bq
\label{eq: seed_integral_motivation}
 {\mathcal J}_1 & = & \frac{N(\eps)}{\omega(x)} \integralseed.
\eq
The function $N(\eps)$, which only depends on $\eps$ but not on $x$, is not essential here. 
It is usually just a power of $\eps$ and in that case 
adjusts the order at which the $\eps$-expansion of ${\mathcal J}_1$ starts.
It is clear from the product rule of differentiation, that if the Picard--Fuchs operator of $\integralseed$ has an
$\eps$-dependent apparent singularity, so will the Picard--Fuchs operator for ${\mathcal J}_1$ and vice versa.

To summarize, we are interested in cases where either no basis exists such that at least one master integral has a Picard--Fuchs operator free of $\eps$-dependent apparent singularities, or where we are not able to find such a basis.
The Feynman integral discussed in this paper falls exactly into this category.
Our aim is to develop a systematic approach to construct a basis for an $\eps$-factorized differential equation for these cases.

\section{A Calabi--Yau Feynman integral for classical gravity} 
\label{sec:analysis_CY}

The focus of this paper is the particular four-loop Feynman integral depicted in fig.~\ref{fig: diagrams}(d), with figs.~\ref{fig: diagrams}(a)--(c) being its three relevant subsectors. In section~\ref{sec:PF_CY}, we show that the associated Picard--Fuchs operator indeed possesses an $\varepsilon$-dependent apparent singularity, such that its differential equation cannot be brought into canonical form using the techniques in refs.~\cite{Pogel:2022ken,Pogel:2022vat}. Moreover, in section~\ref{sec:CY_operator_4dim}, we characterize the geometry encapsulated in the integral in four dimensions. We show that the associated Picard--Fuchs operator satisfies the conditions for a Calabi--Yau operator, proving that the geometry is a Calabi--Yau three-variety.

The Feynman integral in fig.~\ref{fig: diagrams}(d) occurs in the study of the classical dynamics and gravitational waves emitted during the inspiral of two compact astrophysical objects bound in a binary system \cite{Frellesvig:2023bbf}. Specifically, the effective one-body formalism and analytic continuation relate the bound system to the scattering problem~\cite{Damour:2016gwp,Kalin:2019rwq,Kalin:2019inp,Cho:2021arx,Dlapa:2024cje}, allowing us to calculate the classical corrections to the dynamics via the modern scattering-amplitudes approach \cite{Bern:2019crd,Kalin:2020mvi,Mogull:2020sak}. In particular, they are computed by the $2\!\to\!2\,$ scattering of massive scalars, which model the black holes under the assumption of long-range interactions via graviton exchanges. Then, a diagram at $L$ loops is of order $G^{L+1}$, corresponding to the $(L+1)$PM correction in the expansion. In the diagrams, we denote the massive scalars by thick lines, while the thin lines denote the graviton propagators. 

In order to extract the classical pieces from the result, one needs to perform the so-called soft expansion~\cite{Neill:2013wsa}, where the momentum transfer $q$ between the two black holes is small, $|q| \ll 1$. In this classical limit, the propagators for the massive scalars become linearized and take the form $1/(2m_i \, u_i \cdot k_j)$, with $k_j$ being the loop momentum and where $m_i$ and $u_i$ are the mass and normalized four-momenta of the two black holes, respectively, satisfying $u_i^2=1$ and $u_i \cdot q=0$. Thus, the mass dependence totally factors out, and under the soft expansion the scattering process depends only on one dimensionless parameter $y=u_1 \cdot u_2$ and a scale $q^2$ that can be set to $q^2=-1$ and recovered via dimensional analysis. The dimensionless parameter is commonly rewritten as $y=(1+x^2)/2x$ to rationalize the square root $\sqrt{y^2-1}$ which regularly appears in the results. Finally, under the soft expansion, classical diagrams contain at least one massive scalar propagator per loop, and no direct contact between the scalars is allowed. In our case, the classical subsectors of fig.~\ref{fig: diagrams}(d) are given by figs.~\ref{fig: diagrams}(a)--(c). For further details on how Feynman integrals arise in the PM expansion, see refs.~\cite{Bjerrum-Bohr:2022blt,Buonanno:2022pgc}.

\subsection{Picard--Fuchs operator and \texorpdfstring{$\varepsilon$}{epsilon}-dependent apparent singularity}
\label{sec:PF_CY}

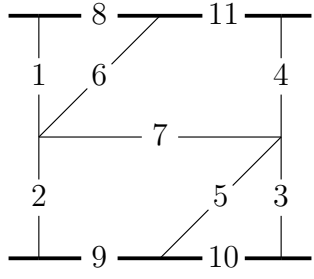
\begin{figure}[t]
\centering
\begin{tikzpicture}[baseline=(current bounding box.center), scale=0.8] 
	\node[] (a) at (0,0) {};
	\node[] (a1) at (2,0) {};
	\node[] (a2) at (4,0) {};
	\node[] (b) at (0,-2) {};
	\node[] (b1) at (4,-2) {};
	\node[] (c) at (0,-4) {};
	\node[] (c1) at (2,-4) {};
	\node[] (c2) at (4,-4) {};
	\node
	 (p1) at ($(a)+(-0.5,0)$) {};
	\node
	 (p2) at ($(c)+(-0.5,0)$) {};
	\node
	 (p3) at ($(c2)+(0.5,0)$) {};
	\node
	(p4) at ($(a2)+(0.5,0)$) {};
	\draw[line width=0.15mm] (b.center) -- (0,-1.3);
	\node[label={[xshift=0cm, yshift=-0.4cm]$1$}] (l1) at (0,-1) {};
	\draw[line width=0.15mm] (0,-0.7) -- (a.center);
	\draw[line width=0.15mm] (b.center) -- (0,-2.7);
	\node[label={[xshift=0cm, yshift=-0.4cm]$2$}] (l2) at (0,-3) {};
	\draw[line width=0.15mm] (0,-3.3) -- (c.center);
	\draw[line width=0.15mm] (b1.center) -- (4,-2.7);
	\node[label={[xshift=0cm, yshift=-0.4cm]$3$}] (l3) at (4,-3) {};
	\draw[line width=0.15mm] (4,-3.3) -- (c2.center);
	\draw[line width=0.15mm] (b1.center) -- (4,-1.3);
	\node[label={[xshift=0cm, yshift=-0.4cm]$4$}] (l4) at (4,-1) {};
	\draw[line width=0.15mm] (4,-0.7) -- (a2.center);
	\draw[line width=0.15mm] (b1.center) -- (3.2,-2.8);
	\node[label={[xshift=0cm, yshift=-0.4cm]$5$}] (l5) at (3,-3) {};
	\draw[line width=0.15mm] (2.8,-3.2) -- (c1.center);
	\draw[line width=0.15mm] (a1.center) -- (1.2,-0.8);
	\node[label={[xshift=0cm, yshift=-0.4cm]$6$}] (l6) at (1,-1) {};
	\draw[line width=0.15mm] (0.8,-1.2) -- (b.center);
	\draw[line width=0.15mm] (b.center) -- (1.7,-2);
	\node[label={[xshift=0cm, yshift=-0.4cm]$7$}] (l7) at (2,-2) {};
	\draw[line width=0.15mm] (2.3,-2) -- (b1.center);
	\draw[line width=0.5mm] (a.center) -- (0.7,0);
	\node[label={[xshift=0cm, yshift=-0.4cm]$8$}] (l8) at (1,0) {};
	\draw[line width=0.5mm] (1.3,0) -- (a1.center);
	\draw[line width=0.5mm] (c.center) -- (0.7,-4);
	\node[label={[xshift=0cm, yshift=-0.4cm]$9$}] (l9) at (1,-4) {};
	\draw[line width=0.5mm] (1.3,-4) -- (c1.center);
	\draw[line width=0.5mm] (c1.center) -- (2.7,-4);
	\node[label={[xshift=0cm, yshift=-0.4cm]$10$}] (l10) at (3.05,-4) {};
	\draw[line width=0.5mm] (3.4,-4) -- (c2.center);
	\draw[line width=0.5mm] (a1.center) -- (2.7,0);
	\node[label={[xshift=0cm, yshift=-0.4cm]$11$}] (l11) at (3.05,0) {};
	\draw[line width=0.5mm] (3.4,0) -- (a2.center);
	\draw[line width=0.5mm] (p1.center) -- (a.center);
	\draw[line width=0.5mm] (a2.center) -- (p4.center);
	\draw[line width=0.5mm] (p2.center) -- (c.center);
	\draw[line width=0.5mm] (c2.center) -- (p3.center);
\end{tikzpicture}
\caption{Convention for the propagators for the integral family of the diagram in fig.~\ref{fig: diagrams}(d), which gives rise to the Calabi--Yau three-variety.}
\label{fig: diag_propagators_CY}
\end{figure}

Let us now look in detail at the four-loop integral of fig.~\ref{fig: diagrams}(d), using the conventions specified in fig.~\ref{fig: diag_propagators_CY}. We consider a family of integrals of the form
\begin{equation}
\label{eq: Calabi--Yau integral family}
\mathcal{I}_{\nu_1\nu_2\dots\nu_{22}}=\int \frac{d^D k_1 \, d^D k_2 \, d^D k_3 \, d^D k_4}{\rho_1^{\nu_1} \rho_2^{\nu_2} \cdots \rho_{22}^{\nu_{22}}},
\end{equation}
where the dependence on the masses, which factor out, has been removed. The 11 propagators are defined as
\begin{align}
\rho_1&=k_1^2, & \rho_2&=k_3^2, & \rho_3&=(k_4{+}q)^2, & \rho_4&=(k_2{-}q)^2, \nonumber \\[0.2cm]
\rho_5&=(k_4{-}k_3)^2, & \rho_6&=(k_1{-}k_2)^2, & \rho_7&=(k_2{+}k_3)^2, & & \\[0.2cm]
\rho_8&=2u_1 \cdot k_1, & \rho_9&=2u_2 \cdot k_3, & \rho_{10}&=2u_2 \cdot k_4, & \rho_{11}&=2u_1 \cdot k_2, \nonumber
\end{align}
and the 11 irreducible scalar products (ISPs) are
\begin{align}
\rho_{12}&=k_2^2, & \rho_{13}&=k_4^2, & \rho_{14}&=(k_1{-}q)^2, & \rho_{15}&=(k_3{+}q)^2, \nonumber \\[0.2cm]
\rho_{16}&=(k_1{+}k_3)^2, & \rho_{17}&=(k_1{+}k_4)^2, & \rho_{18}&=(k_2{+}k_4)^2, & & \\[0.2cm]
\rho_{19}&=2u_2 \cdot k_1, & \rho_{20}&=2u_2 \cdot k_2, & \rho_{21}&=2u_1 \cdot k_3, & \rho_{22}&=2u_1 \cdot k_4. \nonumber
\end{align}

As can be seen, the propagators $\rho_8,\dots,\rho_{11}$ and the ISPs $\rho_{19},\dots,\rho_{22}$ are odd under the so-called parity transformation $u_i \to - u_i$, while the remaining ones are even. Therefore, sectors where the sum of the powers $\nu_8 + \cdots + \nu_{11} + \nu_{19} + \cdots + \nu_{22}$ is even or odd behave differently under IBPs and differential equations, and decouple from each other. In our particular four-loop case, the odd-parity integrals can be written in terms of multiple polylogarithms; hence, we will focus on the even sector, which contains the Calabi--Yau integral.

Using the Baikov representation, in ref.~\cite{Frellesvig:2023bbf} it was found that the leading singularity (LS) of the scalar integral in $D=4$ dimensions takes the form
\begin{equation}
\label{eq: LS_diag_Calabi_Yau}
\LS \big( \mathcal{I}_{\underbrace{\scriptstyle 1\ldots1}_{11}\underbrace{\scriptstyle  0\ldots0}_{11}} \big) \propto  x \int \frac{d t_1 d t_2 d t_3}{\sqrt{P_8(t_1,t_2,t_3)}}\,,
\end{equation}
where $P_8(t_1,t_2,t_3)$ is a polynomial of degree 8 in three variables. Homogenizing the polynomial as $\widetilde{P}_8(t_1,t_2,t_3,t_4)=P_8(t_1/t_4,t_2/t_4,t_3/t_4)\, t_4^8$, the equation
\begin{equation}
 t_5^2-\widetilde{P}_8(t_1,t_2,t_3,t_4)=0
\end{equation}
defines a Calabi--Yau three-variety in weighted projective space $[t_1,t_2,t_3,t_4,t_5]\sim [\lambda^1 t_1,\lambda^1 t_2, \lambda^1 t_3, \lambda^1 t_4,\lambda^4 t_5]\in\mathbb{WP}^{1,1,1,1,4}$~\cite{Hubsch:1992nu,Bourjaily:2019hmc}.

From eq.~\eqref{eq: LS_diag_Calabi_Yau}, the scalar integral looks like a reasonable seed integral $\integralseed$ for its sector, since its leading singularity is an integral over the unique nowhere-vanishing holomorphic top-form; hence, it is a period of the Calabi--Yau three-variety and a candidate for $\omega(x)$ in eq.~\eqref{eq: seed_integral_motivation}. It will play a role in bringing the differential equation into $\varepsilon$-factorized form, as briefly introduced in eq.~\eqref{eq: seed_integral_motivation} and detailed in section~\ref{sec:eps-form_general}.

We can also study the corresponding Picard--Fuchs operator, which in dimensional regularization $D=4-2\varepsilon$ takes the form
\begin{equation}
\label{eq:PF_L5}
\mathcal{L}_5 = \frac{d^5}{d x^5} + \sum_{j=0}^{4} \frac{C_j(x,\varepsilon)}{C_5(x,\varepsilon)} \frac{d^j}{d x^j}\,,
\end{equation}
where $C_j(x,\varepsilon)$ are rational polynomials in $x$ and $\varepsilon$, given in appendix~\ref{app:PF_coefficients}, with the common denominator 
\begin{equation}
C_5(x,\varepsilon) = x^5 (1-x^2)^4 P_{\text{apparent}}(x,\varepsilon).
\end{equation}
As can be seen, the operator contains singularities at $x=0$ and $x=\pm 1$, which correspond to the high-energy limit ($y \to \infty$) and the static limit ($y \to \pm 1$), respectively. Moreover, there is also a singularity at $P_{\text{apparent}}(x,\varepsilon)=0$, with
\begin{equation}
\label{apparent_singularity}
P_{\text{apparent}}(x,\varepsilon) \equiv (3 + 32 \varepsilon + 4 \varepsilon^2) (1 + x^2)^2 - 
 16 x^2 (1 + 5 \varepsilon) (1 + 10 \varepsilon)
\end{equation}
being quartic in $x$ and quadratic in $\varepsilon$. This is however unphysical, and no solutions are expected to be singular at the $\varepsilon$-dependent value for which eq.~\eqref{apparent_singularity} vanishes. Therefore, the Picard--Fuchs operator contains an $\varepsilon$-dependent apparent singularity, which impedes the use of the ansatz in ref.~\cite{Pogel:2022vat} to bring the corresponding differential equation into $\varepsilon$-factorized form.

At this point, we also tried a number of methods available in the literature to remove the $\varepsilon$-dependent apparent singularity from the Picard--Fuchs operator. First, there exist different desingularization methods~\cite{TSAI2000747,Abramov:2004,CHEN2016617,10.1016/j.jsc.2019.02.009}, with computer implementations in \texttt{Maple}~\cite{Abramov:2004} and \texttt{SageMath}~\cite{Kauers:2013juk}. However, these methods both generate further apparent singularities containing polynomials of higher order in $\varepsilon$, and yield a differential operator of higher order than the original, the solutions of which are not totally independent. Similarly, there also exist methods to generate apparent singularities by computing derivatives of a regular operator~\cite{Slavyanov:2016}, yet the inverse operation is not systematic and fails to produce a primitive operator without the $\varepsilon$-dependent apparent singularity starting from $\mathcal{L}_5$. 
Lastly, we also performed an extensive search for potential basis integrals within the family~\eqref{eq: Calabi--Yau integral family}, aiming to identify a candidate free from $\varepsilon$-apparent singularities in the Picard--Fuchs operator. This was done by considering combinations of dots\footnote{As usual, the power of a propagator minus one is referred to as the number of dots on this propagator.} and numerators, 
with the sum of the number of dots and the number of numerators being $\leq 3$. 
There are 1220 such candidates in the even-parity sector. 
We then performed the reduction onto the master integrals of the maximal cut
and considered the subset for which the results of the reductions were different up to an overall 
factor constant in $x$ (but allowed to be a function of $\eps$). 
This reduced the number of candidates to 435, but all of them had a Picard--Fuchs operator with an $\eps$-dependent apparent singularity. In principle, we can extend this search to linear combinations of integrals with 
possibly $x$-dependent coefficients.
This is computationally more expensive and for the few cases we tried we did not find any linear combination for which the $\eps$-dependent apparent singularity disappeared.

\subsection{A Calabi--Yau operator in four dimensions}
\label{sec:CY_operator_4dim}

Before developing a method that can bring the differential equation into $\varepsilon$-factorized form for cases with $\varepsilon$-dependent apparent singularities, let us first study the operator for $\varepsilon=0$, showing that it is indeed a Calabi--Yau operator as originally announced in ref.~\cite{Frellesvig:2023bbf}. In $D=4$ dimensions, the Picard--Fuchs operator~\eqref{eq:PF_L5} factorizes into a product of two lower-order operators $\mathcal{L}_5^{(0)} \equiv \mathcal{L}_5\big|_{\varepsilon=0} = \mathcal{L}_1 \cdot \mathcal{L}_4$, with
\begin{align}
\label{eq: PF_L1_4d}
\mathcal{L}_1 = &\, 
 \frac{d}{d x} + \frac{3 + 4 x^2 - 22 x^4 + 84 x^6 - 21 x^8}{x (1 - x^4)(3-x^2)(1-3x^2)}\,, \\[0.2cm]
\mathcal{L}_4 = &\, 
 \frac{d^4}{d x^4} + \frac{2 - 16 x^2 - 10 x^4}{x(1-x^4)} \frac{d^3}{d x^3} + \frac{1 - 28 x^2 + 46 x^4 + 68 x^6 + 25 x^8}{x^2 (1 - x^4)^2} \frac{d^2}{d x^2} \nonumber \\
& \, - \frac{1 + 11 x^2 - 54 x^4 + 22 x^6 + 37 x^8 + 15 x^{10}}{x^3 (1 - x^2)^3 (1 + x^2)^2} \frac{d}{d x} + \frac{1 + 3 x^2 + 20 x^4 + 3 x^6 + x^8}{x^4 (1 - x^4)^2}\,.
\label{eq: PF_L4_4d}
\end{align}
Since the operator $\mathcal{L}_4$ is irreducible, it indicates the presence of a non-trivial geometry in the integral in $D=4$, which in turn is compatible with a Calabi--Yau three-variety, as identified in eq.~\eqref{eq: LS_diag_Calabi_Yau} from a leading singularity perspective~\cite{Frellesvig:2023bbf}.

To corroborate that we indeed have a Calabi--Yau operator, we can verify that a set of defining properties are satisfied for the operator $\mathcal{L}_4$. We follow the classification of ref.~\cite{Bogner:2013kvr}, which requires the following properties for an operator $\mathcal{L}$ in the variable $x$ to be of Calabi--Yau type:
\begin{enumerate}
	\item The operator is self-dual.
	\item All indicials of the solution space at $x=0$ are equal and integral, i.e.\ $x=0$ is a point of maximal unipotent monodromy (MUM-point).
	\item The holomorphic solution at $x=0$ has an $N$-integral series representation.
	\item The special coordinate (corresponding to the mirror map) has an $N$-integral series representation at the MUM-point $x=0$.
	\item The structure series of the operator have an $N$-integral series representation at the MUM-point $x=0$.
\end{enumerate}
Let us explicitly verify that the operator $\mathcal{L}_4$ satisfies these conditions. 
We will proceed step-by-step, introducing relevant concepts when necessary for the discussion.
Unless stated otherwise, we assume that
\begin{align}
    \mathcal{L}_n &= \frac{\diff^n}{\diff x^n} + \sum_{i=0}^{n-1} a_i(x)\frac{\diff^i}{\diff x^i}\,,
\end{align}
which is related to the notation in eq.~\eqref{eq:PF_L5} by $a_i \equiv C_i/C_n$, but now for $\varepsilon=0$.
Furthermore, we denote the set of roots of $C_n(x)$ to be the set of singularities $S_{\mathcal{L}_n}$ of the operator.

We begin with the first condition, self-duality.
Associate to an operator $\mathcal{L}_n$ the dual operator
\begin{equation}
	\hat{\mathcal{L}}_n=\sum_{i=0}^n (-1)^{n+i}\frac{\diff^i}{\diff x ^i} a_i(x)\,,
\end{equation}
such that the derivatives also act on the coefficients $a_i(x)$.
The operator is then said to be self-dual\footnote{Sometimes the dual operator is called adjoint operator, and the self-dual property is denoted by essential self-adjointness.} if there exists a rational function $\alpha(x) \in \mathbb{Q}(x)$ such that 
\begin{equation}
	\mathcal{L}_n\alpha(x)=\alpha(x) \hat{\mathcal{L}}_n\,.
\end{equation}
Assuming that the operator $\mathcal{L}_n$ is normalized such that $a_n(x)=1$ and comparing the coefficients of $\frac{\diff^{n-1}}{\diff x ^{n-1}}$ in this equation, such a function has to satisfy the simple differential equation
\begin{equation}
	\label{eq:alpha_self-adj}
	\frac{\diff \alpha}{\diff x}=-\frac{2 }{n} a_{n-1}(x)\alpha(x)\,.
\end{equation}
We verify that the operator $\mathcal{L}_4$ at hand is indeed self-dual in this sense, with the function $\alpha(x)$ being (up to a constant normalization)
\begin{equation}
	\label{eq:alpha_L4}
	\alpha(x)=\frac{ x^2+1}{x \left(1-x^2\right)^3}\,.
\end{equation}
For Calabi--Yau operators, the function $\alpha(x)$ is also called the Yukawa coupling.

Next, we turn to the second condition, concerning the indicials at the point $x=0$. For this purpose we need to define the Riemann-$\mathcal{P}$ symbol of an operator $\mathcal{L}_n$.
First, we define the logarithmic derivative $\theta_x=x\frac{\diff}{\diff x}$, which satisfies the identities
\begin{equation}
	\frac{\diff^k}{\diff x^k}=\frac{1}{x^k}\sum_{i=1}^{k}S_1(k,i)\theta_x^i, \qquad \theta_x^k=\sum_{i=1}^{k}S_2(k,i) x^i \frac{\diff^i}{\diff x^i},
\end{equation}
where $S_1$ and $S_2$ are the Stirling numbers of the first and second kind, respectively.
Then, we can rewrite the operator $\mathcal{L}_n$ in the so called $\theta$-form (modulo a normalization in $\mathbb{Q}[x]$)
\begin{equation}
	\mathcal{L}_n\propto P_0 (\theta_x)+x  P_1(\theta_x)+\mathellipsis+x^r  P_r(\theta_x)\,,
\end{equation}
with $P_i\in \mathbb{Q}[\theta]$.
The roots $b_i^{(0)}\in\mathbb{Q}$, $i=\{1,\dots,n\}$, of the polynomial $P_0$ are called the indicials or exponents of the operator $\mathcal{L}_n$ at $x=0$.
To obtain the indicials $b^{(p)}_i$ at arbitrary points $x=p$, the procedure above can be repeated after changing variables to $x_p=x-p$, or $x_\infty=1/x$ in the case of $p=\infty$.
The indicials determine the leading behavior of the solutions of the operator at $x=0$: for an indicial $b_i$ there exists a local series solution (possibly logarithmically divergent) starting at order $x^{b_i}$.
At regular points, the indicials are distinct and increase in integer steps. 
In a neighborhood of such points, the solutions are single-valued. 
Thus, the solution space of the operator is spanned by convergent, holomorphic series solutions starting at $x^{b_1},\mathellipsis,x^{b_n}$.
By contrast, for regular singular points, which satisfy that the coefficients $a_i(x)$ of the operator have a pole of order at most $n-i$, the indicials can be equal. The Riemann-$\mathcal{P}$ symbol is defined as an $n \times k$ matrix collecting the $n$ indicials for the $k$ singular points $p_i\in S_{\mathcal{L}_n}$ of an operator $\mathcal{L}_n$.
Specifically,
\begin{equation}
	\mathcal{P}(\mathcal{L}_n)=\begin{Bmatrix}
		p_1 & \mathellipsis & p_k\\\hline\hline
		b_1^{(p_1)} & \mathellipsis & b_1^{(p_k)}\\
		\vdots & & \vdots\\
		b_n^{(p_1)} & \mathellipsis & b_n^{(p_k)}\\
	\end{Bmatrix}\,.
\end{equation}

Regular singular points for which all indicials are equal are called points of maximal unipotent monodromy (MUM-points).
At these points, the $n$ solutions $\varpi_i$, $i=\{1,\mathellipsis,n\}$ of the operator consist of exactly one holomorphic series solution $\varpi_1$, and $(n-1)$ divergent solutions ascending in powers of $\log(x)$,
taking the shape
\begin{equation}
	\label{eq:Frobenius_MUM}
	\begin{aligned}
		\varpi_1&=\Sigma_1\,,\\
		\varpi_2&=\frac{1}{2\pi i}\left[\log(x)\Sigma_1+\Sigma_2\right]\,,\\[-0.7em]
		&\vdots\\[-1em]
		\varpi_n&=\frac{1}{(2\pi i)^{n-1}}\left[\frac{\log(x)^{n-1}}{(n-1)!}\Sigma_1+\frac{\log(x)^{n-2}}{(n-2)!}\Sigma_2+\mathellipsis+\Sigma_n\right]\,,
	\end{aligned}
\end{equation}
where $\Sigma_i\in\mathbb{Q}[[x]]$.
For the proposed Calabi--Yau operator $\mathcal{L}_4$, we find the set of singularities $S_{\mathcal{L}_4}=\{\pm 1, \pm i, 0, \infty\}$,
and a Riemann-$\mathcal{P}$ symbol
\begin{equation}
\mathcal{P}(\mathcal{L}_4)=\left\{
	\begin{array}{cccccc}
 +1 & -1 & +i & -i & 0 & \infty  \\\hline\hline
 
\begin{array}{c}
 0 \\
 0 \\
 0 \\
 0 \\
\end{array}
 & 
\begin{array}{c}
 0 \\
 0 \\
 0 \\
 0 \\
\end{array}
 & 
\begin{array}{c}
 0 \\
 1 \\
 3 \\
 4 \\
\end{array}
 & 
\begin{array}{c}
 0 \\
 1 \\
 3 \\
 4 \\
\end{array}
 & 
\begin{array}{c}
 1 \\
 1 \\
 1 \\
 1 \\
\end{array}
 & 
\begin{array}{c}
 1 \\
 1 \\
 1 \\
 1 \\
\end{array}
 \\
\end{array}
\right\}\,.
\end{equation}
At this point let us remark on some features. First, there is an evident doubling of singularities due to our choice of variable.
If we make a change of variable to $v=y^2=(1+x^2)^2/4x^2$, the Picard--Fuchs operator~\eqref{eq:PF_L5} can be written as\footnote{The operator in eq.~(17) of ref.~\cite{Klemm:2024wtd} can be further obtained via $z=(1-v)/2^{10}$.}
\begin{equation}
	\mathcal{L}_{4,v}=\frac{\diff^4}{\diff v^4}
	+\frac{2-8 v}{v-v^2}\frac{\diff^3}{\diff v^3}
	+\frac{58 v^2-29 v-1 }{4 (v-1)^2 v^2}\frac{\diff^2}{\diff v^2}
	+\frac{20 v^3-15 v^2-1}{4 (v-1)^3 v^3}\frac{\diff}{\diff v}
   +\frac{4 v^2-v+4}{64 (v-1)^3 v^3},
\end{equation}
with an associated Riemann-$\mathcal{P}$ symbol
\begin{equation}
\mathcal{P}(\mathcal{L}_{4,v})=\left\{\begin{array}{ccc}
 1 & 0 & \infty  \\
\hline\hline
\begin{array}{c}
 0 \\
 0 \\
 0 \\
 0 \\
\end{array}
 & 
\begin{array}{c}
 0 \\
 {1}/ {2} \\
 {3}/ {2} \\
 2 \\
\end{array}
 & 
\begin{array}{c}
 {1}/{2} \\
 {1}/{2} \\
 {1}/{2} \\
 {1}/{2} \\
\end{array}
 \\
\end{array}
\right\}\,.
\end{equation}
While the variable $v$ leads to a more compact operator, we will work with $x$, as this choice rationalizes square roots that appear later on.

Second, we notice that the indicials of the points $x=\pm i$ are integral and distinct. 
While the points $x=\pm i$ constitute a singularity of the coefficients of the operator $\mathcal{L}_4$, the integer steps and the jump in the indicials indicate that the solution space at these points might be just spanned by holomorphic series, similar to a regular, non-singular point.
Indeed, it is easy to check that the Frobenius basis of solutions is made up of holomorphic solutions with series expansions starting at $x^0$, $x^1$, $x^3$, $x^4$.
Thus, the points $x=\pm i$ are only singularities of the operator but not of the solution space.
Such points are called apparent singularities, which we already introduced in section \ref{sec:motivation}.
For Calabi--Yau operators, apparent singularities will arise at the roots of the rational function $\alpha(x)$ by which they are self-dual.
Indeed we see that eq.~\eqref{eq:alpha_L4} vanishes for $x=\pm i$.
In contrast to the aforementioned $\varepsilon$-dependent apparent singularities, these ones are independent of $\varepsilon$ and thus do not pose a problem.%
\footnote{The change of parameters to $v$ introduces a coordinate singularity at this point, which allows to identify the Picard--Fuchs operator in a list of classified Calabi--Yau operators~\cite{Klemm:2024wtd}.}
Lastly, we see that the points $x=\{ \pm1,0,\infty \}$ are MUM-points since all indicials are equal and integer.
Thus, the operator $\mathcal{L}_4$ also satisfies the second condition for a Calabi--Yau operator.

Let us now turn to the third condition. The Frobenius solution of $\mathcal{L}_4$ at $x=0$ has the form of eq.~\eqref{eq:Frobenius_MUM}, with 
series
\begin{equation}
\label{eq:Frobenius series}
	\begin{aligned}
		\Sigma_1&=x+\frac{9 x^3}{2^4}+\frac{1681 x^5}{2^{12}}+\frac{21609 x^7}{2^{16}}+\frac{74805201 x^9}{2^{28}}+\mathcal{O}(x^{11})\,,\\
		\Sigma_2&=\frac{3 x^3}{2^{3}}+\frac{1517
   x^5}{2^{12}}+\frac{22295 x^7}{2^{16}}+\frac{167072733 x^9}{2^{29}}+\mathcal{O}(x^{11})\,,\\
   \Sigma_3&=\frac{5 x^3}{2^5}+\frac{7289 x^5}{2^{15}}+\frac{1123325
   x^7}{2^{19}3^2}+\frac{2042158233 x^9}{2^{33}}+\mathcal{O}(x^{11})\,,\\
   \Sigma_4&=-\frac{5 x^3}{2^5}-\frac{7865 x^5}{2^{16}}-\frac{2448943 x^7}{2^{20}3^3}-\frac{2162890201
   x^9}{2^{35}}+\mathcal{O}(x^{11})\,.
	\end{aligned}
\end{equation}
While the holomorphic solution $\varpi_1=\Sigma_1$ is not integral, it is $N$-integral: 
A series $\sum_{i=0}^{\infty}c_i x^i\in\mathbb{Q}[[x]]$ is said to be $N$-integral if there exists an integer $N\in \mathbb{N}$ such that $\sum_{i=0}^{\infty}c_i (N x)^i\in\mathbb{Z}[[x]]$; that is, the series becomes integral under appropriate integer variable rescaling.
The holomorphic solution $\varpi_1$ of $\mathcal{L}_4$ is $N$-integral with $N=2^4$, thus satisfying the third condition for a Calabi--Yau operator.%
\footnote{We have verified $N$-integrality up to $\mathcal{O}(x^{1000})$ for all series relevant for the conditions satisfied by a Calabi--Yau operator.}

From the Frobenius basis we can also construct the mirror map $q$---in ref.~\cite{Bogner:2013kvr} also called special coordinate---which is the subject of the fourth condition.
With 
\begin{equation}
	\tau:=\frac{\varpi_2}{\varpi_1}=\frac{1}{2\pi i}\left(\log(x)+\frac{\Sigma_2}{\Sigma_1}\right),
\end{equation}
the mirror map at $x=0$ is defined via
\begin{equation}
\label{eq:q-coord}
	q:=\exp(2\pi i \tau)=x \exp\left(\frac{\Sigma_2}{\Sigma_1}\right).
\end{equation}
Its expansion is
\begin{equation}
	q=x+\frac{3 x^3}{2^3}+\frac{941 x^5}{2^{12}}+\frac{5413 x^7}{2^{15}}+\frac{69120717 x^9}{2^{29}}+\frac{452476279 x^{11}}{2^{32}}+\mathcal{O}(x^{12})\,.
\end{equation} 
We again find that this series is $N$-integral with $N=2^4$, satisfying the fourth condition for a Calabi--Yau operator.

The last condition is on the structure series, which are related to a form of the operator $\mathcal{L}_n$ that manifestly annihilates the Frobenius solutions.
Let us define the operators
\begin{equation}
  \begin{aligned}
    \mathcal{N}_1:=\theta_x\frac{1}{\varpi_1},\quad\mathcal{N}_{i+1}:=\theta_x\alpha_{i}\,\mathcal{N}_{i}\,,
  \end{aligned}
\end{equation}
where 
\begin{equation}
  \alpha_i:=\left( \mathcal{N}_i (\varpi_{i+1})\right)^{-1}
  =\left(\theta_x \frac{\mathcal{N}_{i-1}(\varpi_{i+1})}{\mathcal{N}_{i-1}(\varpi_{i})} \right)^{-1}.
\end{equation}
We have, for instance,
\begin{align}
\mathcal{N}_1 (\varpi_1) =& \, \theta_x (1) = 0\,, \\[0.1cm]
\mathcal{N}_2 (\varpi_1) =& \, \theta_x \alpha_1 \mathcal{N}_1 (\varpi_1) = 0\,, \qquad \mathcal{N}_2 (\varpi_2) = \theta_x \underbrace{\alpha_1 \mathcal{N}_1 (\varpi_2)}_{=1} = 0\,, \\
\mathcal{N}_3 (\varpi_1) =& \, \mathcal{N}_3 (\varpi_2) = 0\,, \qquad \qquad \mathcal{N}_3 (\varpi_3) = \theta_x \underbrace{\alpha_2 \mathcal{N}_2 (\varpi_3)}_{=1} = 0\,.
\end{align}
Then, by construction, the operator $\mathcal{N}_{m}$ annihilates all Frobenius 
solutions $\varpi_i$ for $1\le i\le m$,
\begin{equation}
  \mathcal{N}_{m}(\varpi_i)=\theta_x\alpha_{m-1}\theta_x\ldots\theta_x\underbrace{\alpha_{i-1}\mathcal{N}_{i-1}(\varpi_i)}_{=1}=0\,.
\end{equation}
As a consequence, the operator $\mathcal{N}_{n}$ annihilates all Frobenius solutions.
Up to normalization, we must therefore have
\begin{equation}
  \mathcal{L}_n\propto \mathcal{N}_n=\theta_x\alpha_{n-1}\theta_x\alpha_{n-2}\theta_x\ldots\theta_x\alpha_1\theta_x\frac{1}{\varpi_1}\,.
\end{equation}
The functions $\alpha_i$ are called structure series of the operator $\mathcal{L}_n$.
For a Calabi--Yau operator they are symmetric, $\alpha_{i}=\alpha_{n-i}$ for $1\le i\le n-1$, and must have $N$-integral series expansions.
For the operator $\mathcal{L}_4$, we find
\begin{align}
	\alpha_1 &= 1-\frac{3 x^2}{2^2}-\frac{77 x^4}{2^{10}}-\frac{369 x^6}{2^{13}}-\frac{1722877 x^8}{2^{26}}-\frac{1133073 x^{10}}{2^{26}}+\mathcal{O}(x^{11})\,, \nonumber \\
	\alpha_2 &= 1-\frac{11 x^2}{2^3}+\frac{2381 x^4}{2^{11}}-\frac{20723 x^6}{2^{14}}+\frac{161802701 x^8}{2^{27}}-\frac{1336327183 x^{10}}{2^{30}}+\mathcal{O}(x^{11})\,, \nonumber \\
 	\alpha_3 &= 1-\frac{3 x^2}{2^2}-\frac{77 x^4}{2^{10}}-\frac{369 x^6}{2^{13}}-\frac{1722877 x^8}{2^{26}}-\frac{1133073 x^{10}}{2^{26}}+\mathcal{O}(x^{11})\, .
	\end{align}
The operator $\mathcal{L}_4$ is then
\begin{equation}
	\mathcal{L}_4=\beta\ \theta_x \alpha_3 \theta_x \alpha_2 \theta_x \alpha_1 \theta_x \frac{1}{\varpi_1}\,,
\end{equation}
with $\beta$ being an appropriate normalization factor. The structure series satisfy the required symmetry $\alpha_1=\alpha_3$, and are $N$-integral with $N=2^4$, such that $\mathcal{L}_4$ also satisfies the last condition for a Calabi--Yau operator.
We can thus conclude that the operator $\mathcal{L}_4$ is of Calabi--Yau type.

The structure series are furthermore related to the function $\alpha(x)$ defined in eq.~\eqref{eq:alpha_self-adj}.
For a self-dual operator, one finds~\cite{Bogner:2013kvr}
\begin{equation}
	\prod_{i=1}^{n-1}\frac{\alpha_1}{\alpha_i}=c \frac{(x\alpha_1)^{n-1}\alpha}{\varpi_1^2}\equiv  c K(x)\,,
\end{equation}
where $K(x)$ is the gauge-fixed or normalized Yukawa coupling~\cite{Candelas:1990rm,Morrison:1991cd}, and $c$ is a constant.
The ratios
\begin{equation}
	Y_i=\frac{\alpha_1}{\alpha_i}\,,
\end{equation}
are called $Y$-invariants of a Calabi--Yau operator.
Making the change of variable from $x$ to $q$ as defined in eq.~\eqref{eq:q-coord}, one finds
\begin{equation}
	\mathcal{L}_4\propto \theta_q^2 \frac{1}{Y_2} \theta_q^2\frac{1}{\varpi_1}\,,
\end{equation}
with 
\begin{equation}
	\label{eq:Y2}
	Y_2=1+\frac{5 x^2}{2^3}-\frac{775 x^4}{2^{11}}-\frac{445 x^6}{2^{14}}-\frac{5110375 x^8}{2^{27}}-\frac{27054575 x^{10}}{2^{30}}+\mathcal{O}(x^{12})\,,
\end{equation}
which is again $N$-integral with $N=2^4$.
Note that for Calabi--Yau operators of Calabi--Yau three-folds we have $K(x)=Y_2(x)$.
Furthermore, ref.~\cite{Bogner:2013kvr} showed that a Calabi--Yau operator is the symmetric power of an operator of a Calabi--Yau one-fold (i.e.~an elliptic curve) if and only if $Y_i=1$ for all $1\le i \le n-1$.
Since this is not satisfied in our case, we cannot express the solutions of $\mathcal{L}_4$ in terms of products of elliptic functions, as was for example the case in the three-loop equal-mass banana integral~\cite{Primo:2017ipr} or various other PM integrals associated to K3 geometries~\cite{Bern:2021dqo,Dlapa:2021npj,Ruf:2021egk,Dlapa:2022wdu,Klemm:2024wtd}.%
\footnote{Ref.~\cite{Klemm:2024wtd} related the Calabi--Yau operator $\mathcal{L}_4$ to a Hadamard product of elliptic integrals; this identification, however, does not have an immediate impact on the evaluation of the integral.}

While we have shown that our operator is of Calabi--Yau type as defined in ref.~\cite{Bogner:2013kvr}, it can in some contexts be beneficial to require slightly stronger conditions, as done in ref.~\cite{Klemm:2024wtd} for this particular operator.
Another commonly used definition for Calabi--Yau operators~\cite{vanStraten:2017} is to demand integrality instead of $N$-integrality for conditions 2--4, and instead of $N$-integral structure series, to require integral instanton numbers $n_j$.
These are defined as the coefficients of a Lambert series representation of the gauge-fixed Yukawa coupling in the variable $q$,
\begin{equation}
	K(x(q))=1+\sum_{j=1}^{\infty}\frac{n_j j^3 q^j}{1-q^j}\,.
\end{equation}
It is easy to verify that these slightly stronger conditions are also satisfied for the operator $\mathcal{L}_4$ after rescaling $x$ by $2^4$.
The first 10 instanton numbers are then
\begin{equation}
	\begin{tabular}{c|cccccccccc}
		$j$& 1 & 2 & 3 & 4 & 5 & 6& 7& 8& 9& 10\\ \hline
		$n_j$&0 & 20 & 0 &$-870$ & 0 & 67460& 0 & $-6821070$& 0 & 800369820
	\end{tabular}\,.
\end{equation}

\section{An \texorpdfstring{$\varepsilon$}{epsilon}-factorized form for \texorpdfstring{$\varepsilon$}{epsilon}-dependent apparent singularities} 
\label{sec:eps-form_general}

In this section, we investigate the relation between $\eps$-dependent apparent singularities 
and an $\eps$-factorized differential equation.
We focus on the general picture here, while applying it to the Calabi--Yau integral family \eqref{eq: Calabi--Yau integral family} in the next section.

The essential complication occurs on the maximal cut and hence the problem can be studied by restricting to the maximal cut.
We consider a sector with $n$ master integrals depending on one kinematic variable $x$.
Let us denote the non-$\eps$-factorized differential equation on the maximal cut by
\bq
\label{non_eps_factorized_differential_equation}
 \frac{d}{dx} \mathcal{I} & = & \tilde{A}_x\left(x,\eps\right) \mathcal{I},
\eq
where $\mathcal{I}=(\mathcal{I}_1,\dots,\mathcal{I}_{\NF})^T$ is a vector of the $\NF$ master integrals.
The entries of the $\NF \times \NF$-matrix $\tilde{A}_x$ are rational functions of $x$ and $\eps$.
We further assume that
\bq
\label{derivative_basis_I_1}
 \mathcal{I}_1, \frac{d}{dx} \mathcal{I}_1, \dots, \left( \frac{d}{dx}\right)^{\NF-1} \mathcal{I}_1
\eq
span for generic $x$ this vectors space as well, otherwise the problem would split into smaller blocks~\cite{Adams:2017tga}.
With this assumption we may always convert the system of $\NF$ first-order differential equations to
a single differential equation of order $\NF$ for $\mathcal{I}_1$ and vice versa.
This can be done by transforming the differential equation in eq.~(\ref{non_eps_factorized_differential_equation}) from the basis ${\mathcal I}$ to the basis of eq.~(\ref{derivative_basis_I_1}).
We obtain
\bq
 {\mathcal L}_\NF \; \mathcal{I}_1 & = & 0\,,
\eq
where ${\mathcal L}_\NF$ is the Picard--Fuchs operator.
By clearing denominators we may write ${\mathcal L}_\NF$ in the form
\bq
\label{def_Picard_Fuchs_polynomial}
 {\mathcal L}_\NF
 & = &
 \sum\limits_{j=0}^{\NF} C_j\left(x,\eps\right) \frac{d^j}{dx^j}\,,
\eq
where the $C_j$'s are polynomials in $x$ and $\eps$.
We have an $\eps$-dependent apparent singularity if the coefficient $C_{\NF}(x,\eps)$ of the 
highest derivative does not factorize into a polynomial depending only on $x$ and
a polynomial depending only on $\eps$.
This is the case we are concerned with. 

Our aim is to find a transformation by an $\NF \times \NF$-matrix $U(x,\eps)$ to a new basis ${\mathcal J}$ 
\bq
 {\mathcal J} & = & U {\mathcal I}\,,
\eq
such that 
\bq
\label{eps_factorized_differential_equation}
 d \mathcal{J} & = & \eps A\left(x\right) \mathcal{J}
\eq
is in $\eps$-factorized form.
Here, $A(x)$ is an $\varepsilon$-independent $\NF \times \NF$-matrix whose entries are differential one-forms.
As we have only one kinematic variable $x$, we may write
\bq
 A & = & A_x dx.
\eq
In anticipation of a change of variables from $x$ to a (yet unspecified) variable $\tau$, we denote the Jacobian of this transformation by
\bq 
 J & = & \frac{dx}{d\tau}\,.
\eq
We then have
\bq
\label{eps_factorized_differential_equation_tau}
 A \; = \; A_\tau d\tau\,, & \qquad \mbox{with} \qquad & A_\tau \; = \; A_x J\,,
\eq
and
\bq
 J \frac{d}{dx} \mathcal{J} & = & \eps A_\tau \mathcal{J}\,.
\eq 

Although not required, it is often extremely helpful to assume 
that it is possible to choose the basis $\mathcal{J}$ such that the matrix $A$ (or equivalently $A_x$ or $A_\tau$)
satisfies self-duality \cite{Pogel:2024sdi}.
Self-duality is the statement that 
\bq
 A_{ij} & = & A_{(\NF+1-j)(\NF+1-i)}\,,
\eq
i.e.\ the matrix $A$ is symmetric with respect to the anti-diagonal.
The assumption of self-duality simplifies considerably intermediate calculations.
For our gravity integral, 
we will see a posteriori that this assumption is justified.
This provides further evidence that self-duality might hold for all Feynman integrals; see also ref.~\cite{Duhr:2024xsy} for a recent study.

Let ${\mathcal L}_\NF^{(0)}$ be the $\eps^0$-part of the Picard--Fuchs operator ${\mathcal L}_\NF$ defined in eq.~(\ref{def_Picard_Fuchs_polynomial}),
\bq
 {\mathcal L}_\NF^{(0)} & = &  \sum\limits_{j=0}^{\NF} C_j\left(x,0\right) \frac{d^j}{dx^j}\,,
\eq
and $\omega(x)$ a solution of
\bq
 {\mathcal L}_\NF^{(0)} \omega(x) & = & 0\,,
\eq
i.e.\ a leading singularity of the corresponding integral.
As we briefly introduced in eq.~\eqref{eq: seed_integral_motivation}, the starting point of our ansatz will be to take
\bq
\label{relation_seed_integral}
 {\mathcal J}_1 & = & \frac{N(\eps)}{\omega(x)} \integralseed \,,
\eq
where $\integralseed$ is a Feynman integral from this sector, which we will call the seed integral.
The function $N(\eps)$, which only depends on $\eps$ but not on $x$, is not essential here. 
It is usually just a power of $\eps$ and in that case 
adjusts the order at which the $\eps$-expansion of ${\mathcal J}_1$ starts.
The relevant part is the normalization by $1/\omega(x)$.
It is important to note that $\omega(x)$ depends on $x$ but not on $\eps$.
It is easily checked that the Picard--Fuchs operators of $\integralseed$ and ${\mathcal J}_1$
have (apart from possible trivial prefactors) the same leading coefficient $C_{\NF}(x,\eps)$.
Thus, if the seed integral $\integralseed$ has an $\eps$-dependent apparent singularity, so does ${\mathcal J}_1$.

Following the discussion of section~\ref{sec:motivation} and our findings in section~\ref{sec:PF_CY}
for the particular Calabi--Yau integral family \eqref{eq: Calabi--Yau integral family} in gravity, we do not attempt to answer the question whether or not there exists a seed integral for which the corresponding Picard--Fuchs operator 
would be free of this $\eps$-dependent apparent singularity. 
Instead, we directly proceed to the question whether such integrals admit an $\eps$-factorized differential equation. As we can freely change between the system of first-order differential equations and the Picard--Fuchs operator,
we can first study which $\eps$-factorized differential equations give rise to an $\eps$-dependent apparent singularity in the associated Picard--Fuchs operator.

The case where the matrix $A_\tau$ in eq.~(\ref{eps_factorized_differential_equation_tau}) is lower triangular
\bq
 A_\tau & = &
 \left( \begin{array}{ccccc}
  F_{11} & 0 & 0 & \hdots & 0 \\
  F_{21} & F_{22} & 0 & \hdots & 0 \\
  \vdots & \vdots & \ddots & \ddots & \vdots \\
  F_{(\NF-1) 1} & F_{(\NF-1) 1} & \hdots & F_{(\NF-1)(\NF-1)} & 0 \\
  F_{\NF 1} & F_{\NF 2} & \hdots & F_{\NF (\NF-1)} & F_{\NF \NF} \\
 \end{array} \right)
\eq
is too simple, as ${\mathcal J}_1$ and its derivatives only span a vector subspace of dimension one.
Here, $A_\tau$ and all the functions $F_{ij}$ may depend on $x$, but not on $\eps$.
In order to get an irreducible Picard--Fuchs operator of order $\NF$, we need at least non-zero entries
on the next-to-diagonal:
\bq
\label{A_tau_banana}
 A_\tau & = &
 \left( \begin{array}{cccccc}
  F_{11} & F_{12} & 0 & 0 & \hdots & 0 \\
  F_{21} & F_{22} & F_{23} & 0 & \hdots & 0\\
  \vdots & \vdots & & \ddots & \ddots & \vdots \\
  F_{(\NF-2) 1} & F_{(\NF-2) 2} & \hdots & F_{(\NF-2) (\NF-2)} & F_{(\NF-2) (\NF-1)}  & 0 \\
  F_{(\NF-1) 1} & F_{(\NF-1) 2} & \hdots & F_{(\NF-1) (\NF-2)} & F_{(\NF-1)(\NF-1)} & F_{(\NF-1)\NF} \\
  F_{\NF 1} & F_{\NF 2} & \hdots & F_{\NF (\NF-2)} & F_{\NF (\NF-1)} & F_{\NF \NF} \\
 \end{array} \right).
\eq
A matrix of this form will give rise to a Picard--Fuchs operator of order $\NF$ with leading coefficient
\bq
 C_\NF(x,\eps)
 & = & 
 J^\NF \prod\limits_{i=1}^{\NF-1} F_{i(i+1)}^{\NF-i}\,,
\eq
which is independent of $\eps$ and hence does not lead to an $\eps$-dependent apparent singularity.
A matrix of the form as in eq.~(\ref{A_tau_banana}) occurs in the equal-mass $l$-loop banana integral \cite{Pogel:2022vat}.

In order to get an $\eps$-dependent apparent singularity, we in addition need to consider non-zero entries on the next-to-next-to-diagonal, i.e.\
the matrix $A_\tau$ is of the form
\bq
\label{NND_ansatz}
 A_\tau & = &
 \left( \begin{array}{cccccc}
  F_{11} & F_{12} & \hspace{0.5cm} F_{13} & \hspace{0.5cm} 0 & \hdots & 0 \\
  F_{21} & F_{22} & \hspace{0.5cm} F_{23} & \hspace{0.5cm} F_{24} & \hdots & 0\\
  \vdots & \vdots & & \ddots & \ddots & \vdots \\
  F_{(\NF-3) 1} & F_{(\NF-3) 2} & \hspace{0.5cm} \hdots & \hspace{0.5cm}F_{(\NF-3) (\NF-2)} & F_{(\NF-3) (\NF-1)}  & 0 \\
  F_{(\NF-2) 1} & F_{(\NF-2) 2} & \hspace{0.5cm} \hdots & \hspace{0.5cm} F_{(\NF-2) (\NF-2)} & F_{(\NF-2) (\NF-1)}  & F_{(\NF-2) \NF} \\
  F_{(\NF-1) 1} & F_{(\NF-1) 2} & \hspace{0.5cm} \hdots & \hspace{0.5cm} F_{(\NF-1) (\NF-2)} & F_{(\NF-1)(\NF-1)} & F_{(\NF-1)\NF} \\
  F_{\NF 1} & F_{\NF 2} & \hspace{0.5cm} \hdots & \hspace{0.5cm} F_{\NF (\NF-2)} & F_{\NF (\NF-1)} & F_{\NF \NF} \\
 \end{array} \right).
\eq
This requires $\NF \ge 3$, with the simplest case given by $\NF=3$.
Assuming self-duality, the matrix $A_\tau$ reads in this case
\bq
 A_\tau & = &
 \left( \begin{array}{ccc}
 F_{11} & F_{12} & F_{13} \\
 F_{21} & F_{22} & F_{12} \\
 F_{31} & F_{21} & F_{11} \\
 \end{array} \right),
\eq
and the leading coefficient of the Picard--Fuchs operator is now linear in $\eps$ and given by
\bq
 C_3(x,\eps)
 & = & 
 J^4 \left( F_{12} \frac{d}{dx} F_{13} - F_{13} \frac{d}{dx} F_{12} \right)
 + J^3 \left[ F_{12}^3 - F_{13}^2 F_{21} + F_{12} F_{13} \left( F_{11} - F_{22} \right)  \right] \eps.
 \nonumber \\
\eq
Of interest to us is the case $\NF=5$ with just one non-zero entry on the next-to-next-to-diagonal.
Assuming self-duality, such a matrix $A_\tau$ has the form
\bq
\label{eq:ansatz_section3}
 A_\tau & = &
 \left( \begin{array}{ccccc}
 F_{11} & F_{12} & 0 & 0 & 0 \\
 F_{21} & F_{22} & F_{23} & F_{24} & 0 \\
 F_{31} & F_{32} & F_{33} & F_{23} & 0 \\
 F_{41} & F_{42} & F_{32} & F_{22} & F_{12} \\
 F_{51} & F_{41} & F_{31} & F_{21} & F_{11} \\
 \end{array} \right).
\eq
This will result in a leading coefficient of the Picard--Fuchs operator which is quadratic in $\eps$, as the one in eq.\ \eqref{apparent_singularity}.

\begin{table}
\centering
\begin{tabular}{cccc}
$\left( \! \begin{smallmatrix} \bullet & \bullet & \bullet\\ \bullet & \bullet & \bullet\\ \bullet & \bullet & \bullet \end{smallmatrix} \! \right) \to 1\,, \!\quad$ &
$\left( \! \begin{smallmatrix} \bullet & \bullet & \bullet & \circ \\ \bullet & \bullet & \bullet & \bullet\\ \bullet & \bullet & \bullet & \bullet\\ \bullet & \bullet & \bullet & \bullet \end{smallmatrix} \! \right) \to 2\,, \!\quad$ &
$\left( \! \begin{smallmatrix} \bullet & \bullet & \bullet & \bullet\\  \bullet & \bullet & \bullet & \bullet\\ \bullet & \bullet & \bullet & \bullet\\ \bullet & \bullet & \bullet & \bullet \end{smallmatrix} \! \right) \to 3\,, \!\quad$ & 
$\left( \! \begin{smallmatrix} \bullet & \bullet & \circ & \circ & \circ \\ \bullet & \bullet & \bullet & \bullet & \circ \\ \bullet & \bullet & \bullet & \bullet & \circ \\ \bullet & \bullet & \bullet & \bullet & \bullet \\ \bullet & \bullet & \bullet & \bullet & \bullet \end{smallmatrix} \! \right) \to 2\,,$ \\[0.6cm]
$\left( \! \begin{smallmatrix} \bullet & \bullet & \bullet & \circ & \circ \\ \bullet & \bullet & \bullet & \bullet & \circ \\ \bullet & \bullet & \bullet & \bullet & \bullet \\ \bullet & \bullet & \bullet & \bullet & \bullet \\ \bullet & \bullet & \bullet & \bullet & \bullet \end{smallmatrix} \! \right) \to 4\,, \!\!\quad$ & 
$\left( \! \begin{smallmatrix} \bullet & \bullet & \bullet & \bullet & \circ \\ \bullet & \bullet & \bullet & \bullet & \bullet \\ \bullet & \bullet & \bullet & \bullet & \bullet \\ \bullet & \bullet & \bullet & \bullet & \bullet \\ \bullet & \bullet & \bullet & \bullet & \bullet \end{smallmatrix} \! \right) \to 5\,, \!\quad$ & 
$\left( \! \begin{smallmatrix} \bullet & \bullet & \bullet & \bullet & \bullet \\ \bullet & \bullet & \bullet & \bullet & \bullet \\ \bullet & \bullet & \bullet & \bullet & \bullet \\ \bullet & \bullet & \bullet & \bullet & \bullet \\ \bullet & \bullet & \bullet & \bullet & \bullet \end{smallmatrix} \! \right) \to 6\,, \!\quad$ & $\left( \! \begin{smallmatrix} \bullet & \bullet & \circ & \circ & \circ & \circ \\ \bullet & \bullet & \bullet & \bullet & \circ & \circ \\ \bullet & \bullet & \bullet & \bullet & \bullet & \circ \\ \bullet & \bullet & \bullet & \bullet & \bullet & \circ \\ \bullet & \bullet & \bullet & \bullet & \bullet & \bullet  \\ \bullet & \bullet & \bullet & \bullet & \bullet & \bullet \end{smallmatrix} \! \right) \to 3\,,$ \\[8mm]
$\left( \! \begin{smallmatrix} \bullet & \bullet & \circ & \circ & \circ & \circ \\ \bullet & \bullet & \bullet & \bullet & \bullet & \circ \\ \bullet & \bullet & \bullet & \bullet & \bullet & \circ \\ \bullet & \bullet & \bullet & \bullet & \bullet & \circ \\ \bullet & \bullet & \bullet & \bullet & \bullet & \bullet  \\ \bullet & \bullet & \bullet & \bullet & \bullet & \bullet \end{smallmatrix} \! \right) \to 5\,, \!\quad$ & 
$\left( \! \begin{smallmatrix} \bullet & \bullet & \bullet & \circ & \circ & \circ \\ \bullet & \bullet & \bullet & \bullet & \circ & \circ \\ \bullet & \bullet & \bullet & \bullet & \bullet & \circ \\ \bullet & \bullet & \bullet & \bullet & \bullet & \bullet \\ \bullet & \bullet & \bullet & \bullet & \bullet & \bullet  \\ \bullet & \bullet & \bullet & \bullet & \bullet & \bullet \end{smallmatrix} \! \right) \to 6\,, \!\quad$ & 
$\left( \! \begin{smallmatrix} \bullet & \bullet & \bullet & \circ & \circ & \circ \\ \bullet & \bullet & \bullet & \bullet & \bullet & \circ \\ \bullet & \bullet & \bullet & \bullet & \bullet  & \circ \\ \bullet & \bullet & \bullet & \bullet & \bullet & \bullet \\ \bullet & \bullet & \bullet & \bullet & \bullet & \bullet  \\ \bullet & \bullet & \bullet & \bullet & \bullet & \bullet \end{smallmatrix} \! \right) \to 6\,, \!\quad$ & $\left( \! \begin{smallmatrix} \bullet & \bullet & \bullet & \bullet & \circ & \circ \\ \bullet & \bullet & \bullet & \bullet & \bullet & \circ \\ \bullet & \bullet & \bullet & \bullet & \bullet & \bullet \\ \bullet & \bullet & \bullet & \bullet & \bullet & \bullet \\ \bullet & \bullet & \bullet & \bullet & \bullet & \bullet  \\ \bullet & \bullet & \bullet & \bullet & \bullet & \bullet \end{smallmatrix} \! \right) \to 8\,,$ \\[8mm]
& $\left( \! \begin{smallmatrix} \bullet & \bullet & \bullet & \bullet & \bullet & \circ \\ \bullet & \bullet & \bullet & \bullet & \bullet & \bullet \\ \bullet & \bullet & \bullet & \bullet & \bullet & \bullet \\ \bullet & \bullet & \bullet & \bullet & \bullet & \bullet \\ \bullet & \bullet & \bullet & \bullet & \bullet & \bullet  \\ \bullet & \bullet & \bullet & \bullet & \bullet & \bullet \end{smallmatrix} \! \right) \to 9\,, \!\quad$ & 
$\left( \! \begin{smallmatrix} \bullet & \bullet & \bullet & \bullet & \bullet & \bullet \\ \bullet & \bullet & \bullet & \bullet & \bullet & \bullet \\ \bullet & \bullet & \bullet & \bullet & \bullet & \bullet \\ \bullet & \bullet & \bullet & \bullet & \bullet & \bullet \\ \bullet & \bullet & \bullet & \bullet & \bullet & \bullet  \\ \bullet & \bullet & \bullet & \bullet & \bullet & \bullet \end{smallmatrix} \! \right) \to 10\,.  \!\enspace$ & 
\end{tabular}
\caption{Degree in $\varepsilon$ of the apparent singularity of the Picard--Fuchs operator, as a function of the size and shape of the $\varepsilon$-factorized, self-dual differential equation matrix~\eqref{NND_ansatz}, up to a $6 \times 6$ size.}
\label{tab:epsilondegree}
\end{table}

We may systematically investigate the structure of these $\varepsilon$-dependent apparent singularities for various sizes and shapes for the differential equation matrix.\footnote{Specifically, we use the ansatz
$(A_x)_{ij} = \varepsilon ( \sum_{k=0}^{\text{max}_{ij}} \! c_{ijk} x^k ) / (x^{{\alpha}_{ij}} (1{-}x^2)^{{\beta}_{ij}})$
where $\alpha_{ij}$ and $\beta_{ij}$ are chosen randomly as either zero or one, $\text{max}_{ij}$ randomly from $0$ to $5$, and $c_{ijk}$ as random integers from $-50$ to $50$. We furthermore impose the self dual property, so $(A_x)_{ij} = (A_x)_{(n+1-j), (n+1-i)}$.}
If all entries in the next-to-next-to-diagonal and above are zero, the associated Picard--Fuchs has no $\varepsilon$-dependent apparent singularity. Otherwise, there is an $\varepsilon$-dependent apparent singularity, with the degree in $\varepsilon$ depending on the exact location of the non-zero entries of the matrix. We list these degrees for all options with $n \leq 6$ in table~\ref{tab:epsilondegree}.

Let us now discuss how to work practically with an ansatz of the form~(\ref{NND_ansatz}).
We assume that we know the non-$\eps$-factorized differential equation~(\ref{non_eps_factorized_differential_equation})
for the basis ${\mathcal I}$. 
Without loss of generality, we may assume that we have chosen a pre-canonical basis ${\mathcal I}$ such that ${\mathcal I}_1=\integralseed$.
We further assume the relation between $\integralseed$ and ${\mathcal J}_1$ given in eq.~(\ref{relation_seed_integral}) and that the differential equation for the basis ${\mathcal J}$ is of the form~(\ref{NND_ansatz}).
We treat $\omega$, $J$ and the $F_{ij}$'s as unknown functions, to be determined from consistency conditions.
We stress that the approach does not require to know $\omega$ and $J$ beforehand. However, feeding in information from the geometry will simplify the determination of the remaining unknown functions.

We first construct the matrix $U$ relating the two bases ${\mathcal I}$ and ${\mathcal J}$,
\bq
 {\mathcal J} & = & U {\mathcal I}\,.
\eq
We do this by introducing an intermediate basis ${\mathcal K}$ given by
\bq
 {\mathcal K}_1 & = & {\mathcal J}_1\,,
 \nonumber \\
 {\mathcal K}_j & = & \frac{J}{\eps} \frac{d}{dx} {\mathcal K}_{j-1},
 \;\;\;\;\;\; 2 \le j \le \NF.
\eq
This intermediate basis has the property that we can relate it to the basis ${\mathcal I}$ and to the basis ${\mathcal J}$.
In detail, we first obtain transformation matrices $U_1$ and $U_2$ such that
\bq
 {\mathcal K} \; = \; U_1 {\mathcal I}\,,
 &\qquad &
 {\mathcal K} \; = \; U_2 {\mathcal J}\,.
\eq
We then have
\bq
\label{def_U}
 {\mathcal J} \; = \; U {\mathcal I},
 & \qquad \mbox{with} \qquad &
 U \; = \; U_2^{-1} U_1.
\eq
The next task is to determine the unknown functions 
of our ansatz, i.e.\ the functions $\omega$, $J$ and the functions $F_{ij}$'s appearing in eq.~(\ref{NND_ansatz}).
We discuss two methods to solve this problem.

Within the first method
we transform the differential equation~(\ref{non_eps_factorized_differential_equation})
with the matrix $U$ from eq.~(\ref{def_U}) to the basis ${\mathcal J}$. This yields
\bq
 \tilde{A}_x' & = & U \tilde{A}_x U^{-1} - U \frac{d}{dx} U^{-1}\,.
\eq
This should be equal to
\bq
 \frac{\eps}{J} A_\tau\,,
\eq
where the Jacobian factor $1/J$ takes care of the change of variables from $\tau$ to $x$.
We can now determine the unknown functions of the ansatz from the condition that
\bq
\label{eq_set_v1}
 U \tilde{A}_x U^{-1} - U \frac{d}{dx} U^{-1} - \frac{\eps}{J} A_\tau & = & 0\,,
\eq
i.e.\ we require that the left-hand side is the null matrix.
The entries of the matrix on the left-hand side will be in general rational functions in
the dimensional regularisation parameter $\eps$.
The denominators may contain apparent singularities like the one given in eq.~(\ref{apparent_singularity}).
This is not a problem.
We can look at the numerators, which are polynomials in $\eps$, and require that all coefficients of the various powers of $\eps$ vanish. This gives us the conditions which determine the unknown functions from the ansatz.

Within the second method, we consider the Picard--Fuchs operator for the integral ${\mathcal J}_1$, derived in two ways:
On the one hand, we may derive the Picard--Fuchs operator for ${\mathcal J}_1$ from
the differential equation corresponding to eq.~(\ref{NND_ansatz}).
Clearing denominators, we may write
\bq
\label{PF_v1}
 \left[ \sum\limits_{k=0}^{\NF} \sum\limits_{j \ge 0} p_{k,j}\left(x\right) \eps^j \frac{d^k}{dx^k} \right]
 {\mathcal J}_1 & = & 0\,.
\eq
On the other hand, we may derive the Picard--Fuchs operator for 
\bq
 {\mathcal J}_1 & = & \frac{N(\eps)}{\omega} \integralseed
\eq
by using the non-$\eps$-factorized differential equation for $\integralseed$ from eq.~(\ref{non_eps_factorized_differential_equation}).
Again, after clearing denominators we obtain
\bq
\label{PF_v2}
 \left[ \sum\limits_{k=0}^{\NF} \sum\limits_{j \ge 0} \tilde{p}_{k,j}\left(x\right) \eps^j \frac{d^k}{dx^k} \right]
 {\mathcal J}_1 & = & 0\,.
\eq
The sums over $j$ in eqs.~(\ref{PF_v1}) and~(\ref{PF_v2}) are finite.
The index of the last non-zero term in the sum over $j$ may depend on $k$.  
By comparing eq.~(\ref{PF_v1}) with eq.~(\ref{PF_v2}), we conclude that there must exist a function $c(x)$ such that
\bq
\label{eq_set_v2}
 \tilde{p}_{k,j}\left(x\right) & = & c\left(x\right) \cdot p_{k,j}\left(x\right)
\eq
for all $0 \le k \le \NF$ and all $0 \le j$.

The set of equations obtained from eqs.~(\ref{eq_set_v1}) and~(\ref{eq_set_v2}) are equivalent.
We find that the set of equations coming from eq.~(\ref{eq_set_v2}) naturally has a triangular structure:
One first solves the equations with $k=\NF$ and works downwards towards $k=0$.
Solving this set of equations is simplified by assuming self-duality from the start and by first obtaining 
$\omega$ and $J$ from the Calabi--Yau geometry.

We want to stress that the choice of multiplying the coefficients $p_{k,j}$ by the function $c(x)$ has practical significance.
An efficient way to solve the constraints of eq.~\eqref{eq_set_v2} is to bring both sides over a common denominator, and solving for equality of the numerators.
If we were to assume $c(x)$ to multiply $\tilde{p}_{k,j}$, we would find a number of solutions that set both $c(x)$ and $p_{k,j}$ simultaneously to zero.
These solutions are obviously undesirable, and make solving the constraints significantly harder.

In the next section we apply the latter method to bring the gravity integral of section \ref{sec:analysis_CY} into canonical form, since comparing the Picard--Fuchs operators allows us to directly match the expression for the $\varepsilon$-dependent apparent singularity.

\section{Application to the gravity integral} 
\label{sec:eps-form_CY}
In this section, we apply the method of section~\ref{sec:eps-form_general} to obtain an $\eps$-factorized differential equation for the 5PM integral family~\eqref{eq: Calabi--Yau integral family} described in section~\ref{sec:analysis_CY}.

We use \texttt{FIRE}~\cite{Smirnov:2023yhb} and \texttt{KIRA}~\cite{Klappert:2020nbg} to derive integration-by-parts relations, and obtain a set of pre-canonical master integrals, belonging to four sectors:
\begin{align*}
		\text{Sector $\sectora{}$ } \Big(\!{\begin{tikzpicture}[baseline={([yshift=-0.1cm]current bounding box.center)}, scale=0.6] 
		\node[] (a) at (0,0) {};
		\node[] (a1) at (0.5,0) {};
		\node[] (a2) at (1,0) {};
		\node[] (b) at (0,-1) {};
		\node[] (b1) at (0.5,-1) {};
		\node[] (b2) at (1,-1) {};
		\node[] (p1) at ($(a)+(-0.2,0)$) {};
		\node[] (p2) at ($(b)+(-0.2,0)$) {};
		\node[] (p3) at ($(b2)+(0.2,0)$) {};
		\node[] (p4) at ($(a2)+(0.2,0)$) {};
		\draw[line width=0.15mm] (b.center) -- (a2.center);
		\draw[line width=0.15mm] (b1.center) -- (a1.center);
		\draw[line width=0.15mm] (b2.center) -- (a.center);
		\draw[line width=0.5mm] (p1.center) -- (p4.center);
		\draw[line width=0.5mm] (p2.center) -- (p3.center);
		\end{tikzpicture}}\!\Big):&\enspace \quad  \integralprec_1=\integralprec_{111 111 011 11 000 000 000 00}\\[0.5em]
		\text{Sector $\sectorb{}$ } \Big(\!{\begin{tikzpicture}[baseline={([yshift=-0.1cm]current bounding box.center)}, scale=0.6] 
		\node[] (a) at (0,0) {};
	\node[] (a1) at (0.5,0) {};
	\node[] (a2) at (1,0) {};
	\node[] (c) at (0,-1) {};
	\node[] (c1) at (0.5,-1) {};
	\node[] (c2) at (1,-1) {};
	\node[] (p1) at ($(a)+(-0.2,0)$) {};
	\node[] (p2) at ($(c)+(-0.2,0)$) {};
	\node[] (p3) at ($(c2)+(0.2,0)$) {};
	\node[] (p4) at ($(a2)+(0.2,0)$) {};
	\draw[line width=0.15mm] (c.center) -- (a.center);
	\draw[line width=0.15mm] (c.center) -- (a1.center);
	\draw[line width=0.15mm] (c2.center) -- (a2.center);
	\draw[line width=0.15mm] (c.center) -- (a2.center);
	\draw[line width=0.15mm] (c1.center) -- (a2.center);
	\draw[line width=0.5mm] (p1.center) -- (p4.center);
	\draw[line width=0.5mm] (p2.center) -- (p3.center);
		\end{tikzpicture}}\!\Big):&\enspace \begin{cases}\integralprec_2=\integralprec_{101 011 111 11 000 000 000 00}\\*
\integralprec_3=\integralprec_{101 012 111 11 000 000 000 00}\\*
\integralprec_4=\integralprec_{101 011 211 11 000 000 000 00}\end{cases}\\[0.5em]
		\text{Sector $\sectorc{}$ } \Big(\!{\begin{tikzpicture}[baseline={([yshift=-0.1cm]current bounding box.center)}, scale=0.6] 
			\node[] (a) at (0,0) {};
		\node[] (a1) at (0.5,0) {};
		\node[] (a2) at (1,0) {};
		\node[] (b) at (1,-0.5) {};	
		\node[] (c) at (0,-1) {};
		\node[] (c1) at (0.5,-1) {};
		\node[] (c2) at (1,-1) {};
		\node[] (p1) at ($(a)+(-0.2,0)$) {};
		\node[] (p2) at ($(c)+(-0.2,0)$) {};
		\node[] (p3) at ($(c2)+(0.2,0)$) {};
		\node[] (p4) at ($(a2)+(0.2,0)$) {};
		\draw[line width=0.15mm] (c.center) -- (a.center);
		\draw[line width=0.15mm] (c.center) -- (a1.center);
		\draw[line width=0.15mm] (c2.center) -- (a2.center);
		\draw[line width=0.15mm] (c.center) -- (b.center);
		\draw[line width=0.15mm] (c1.center) -- (b.center);
		\draw[line width=0.5mm] (p1.center) -- (p4.center);
		\draw[line width=0.5mm] (p2.center) -- (p3.center);
		\end{tikzpicture}}\!\Big):&\enspace \quad  \integralprec_5=\integralprec_{101 111 111 11 000 000 000 00}\\[0.5em]
		\text{Sector $\sectord{}$ } \Big(\!{\begin{tikzpicture}[baseline={([yshift=-0.1cm]current bounding box.center)}, scale=0.6] 
			\node[] (a) at (0,0) {};
	\node[] (a1) at (0.5,0) {};
	\node[] (a2) at (1,0) {};
	\node[] (b) at (0,-0.5) {};
	\node[] (b1) at (1,-0.5) {};
	\node[] (c) at (0,-1) {};
	\node[] (c1) at (0.5,-1) {};
	\node[] (c2) at (1,-1) {};
	\node[] (p1) at ($(a)+(-0.2,0)$) {};
	\node[] (p2) at ($(c)+(-0.2,0)$) {};
	\node[] (p3) at ($(c2)+(0.2,0)$) {};
	\node[] (p4) at ($(a2)+(0.2,0)$) {};
	\draw[line width=0.15mm] (c.center) -- (a.center);
	\draw[line width=0.15mm] (b.center) -- (b1.center);
	\draw[line width=0.15mm] (c2.center) -- (a2.center);
	\draw[line width=0.15mm] (b1.center) -- (c1.center);
	\draw[line width=0.15mm] (b.center) -- (a1.center);
	\draw[line width=0.5mm] (p1.center) -- (p4.center);
	\draw[line width=0.5mm] (p2.center) -- (p3.center);
		\end{tikzpicture}}\!\Big):&\enspace \begin{cases}  \integralprec_6=\integralprec_{111 111 111 11 000 000 000 00}\\*
\integralprec_7=\integralprec_{111 211 111 11 000 000 000 00}\\*
\integralprec_8=\integralprec_{111 112 111 11 000 000 000 00}\\*
\integralprec_9=\integralprec_{111 111 211 11 000 000 000 00}\\*
\integralprec_{10}=\integralprec_{111 111 111 11 ({-}1)00 000 000 00}\end{cases}
	\end{align*}
Sector $\sectord{}$ is the topsector, while sectors $\sectorc{}$, $\sectorb{}$, and $\sectora{}$ are subsectors.

The integrals ${\integralprec}=(\integralprec_1,\dots,\integralprec_{10})^T$ satisfy a differential equation in block-triangular form,
\begin{equation}
\label{eq:A_prec}
	\frac{\diff}{\diff x}{\integralprec}=\tilde{A}_x(x,\eps) \, {\integralprec}=
	\begin{pNiceMatrix}[columns-width = 1em]
\Block[draw=blue,fill=blue!40,rounded-corners]{1-1}{}
0  &  &  &  &   &   & & & &  \\
0 & \Block[draw=blue,fill=blue!40,rounded-corners]{3-3}{} \bullet & \bullet  & \bullet  &   &   &  & & &   \\
0 & \bullet  & \bullet  & \bullet  &   &   &  & & &   \\
0 & \bullet  & \bullet  & \bullet  &  &   &  & & &   \\
0  & \Block[draw=red,fill=red!40,rounded-corners]{1-3}{} \bullet  & \bullet & \bullet  & \Block[draw=blue,fill=blue!40,rounded-corners]{1-1}{}\bullet  &  &  & & &  \\
 \Block[draw=red,fill=red!40,rounded-corners]{5-5}{}\bullet  & \bullet  & \bullet  & \bullet & \bullet &\Block[draw=blue,fill=blue!40,rounded-corners]{5-5}{} \bullet  & \bullet &\bullet &\bullet & \bullet \\
\bullet  & \bullet  & \bullet  & \bullet & \bullet & \bullet  & \bullet &\bullet &\bullet & \bullet \\
\bullet  & \bullet  & \bullet  & \bullet & \bullet & \bullet  & \bullet &\bullet &\bullet & \bullet \\
\bullet  & \bullet  & \bullet  & \bullet & \bullet & \bullet  & \bullet &\bullet &\bullet & \bullet \\
\bullet  & \bullet  & \bullet  & \bullet & \bullet & \bullet  & \bullet &\bullet &\bullet & \bullet 
\end{pNiceMatrix}{\integralprec}\,,
\end{equation}
with the entries of $\tilde{A}_x$ being rational functions in $\eps$ and $x$.

To find a transformation to an $\eps$-factorized form for this differential equation, we work our way from subsectors to the topsector, starting with the maximal cuts (in blue) and afterwards bringing the off-diagonal terms (in red) into $\varepsilon$-factorized form. In particular, we proceed as follows.
We first use the method of section~\ref{sec:eps-form_general} to find a transformation to $\eps$-factorize the diagonal block of the only non-trivial subsector $\sectorb{}$, leading to a matrix consisting of $\diff\log$ forms.
The diagonal block of sector $\sectorc{}$ is already proportional in $\eps$.
We then use \texttt{Fuchsia} \cite{Gituliar:2017vzm} to also bring the off-diagonal block mixing sectors $\sectorb{}$ and $\sectorc{}$ into canonical form.
The full $5\times 5$ block of subsector integrals $\integralprec_1$--$\integralprec_5$ is then $\eps$-factorized and consisting of $\diff\log$ forms.
We then turn to the main challenge of the computation, the diagonal block of the topsector. We use the ansatz shown in eq.~\eqref{eq:ansatz_section3} to bring this block into $\eps$-factorized form, and show how to derive and solve the corresponding constraints. Finally, we refine our set of integrals to also bring the remaining off-diagonal $5\times 5$ block linking the top- and subsectors into $\eps$-factorized form.
In the following, we will denote the integral basis with $\eps$-factorized diagonal blocks by $\integralFinal_i$, and the final refined basis with also $\eps$-factorized off-diagonal blocks by $\integralFinalReally_i$.
We are not concerned by the overall normalization of the final refined basis
$\integralFinalReally$, as multiplying all master integrals from the basis
$\integralFinalReally$ by the same $\eps$-dependent but $x$-independent function $N(\eps)$
will preserve the $\eps$-factorized form.

\subsection{Subsectors}
Let us consider first the diagonal blocks of the subsector integrals.
The $1\times 1$ diagonal blocks of sectors $\sectora{}$ and $\sectorc{}$ are $0$ and $\eps$-factorized, respectively,
such that 
\begin{equation}
	\begin{aligned}
		\integralFinal_1^{(\sectora{})}&=\integralprec_1\,,\\
		\integralFinal_1^{(\sectorc{})}&=\integralprec_5\,.
	\end{aligned}
\end{equation}
We thus only need to consider the $3\times 3$ block of sector $\sectorb{}$.
A leading singularity analysis based on refs.~\cite{Frellesvig:2024zph,Frellesvig:2024ymq} suggests that this sector is purely polylogarithmic, and can therefore be treated with standard tools such as \texttt{Fuchsia}~\cite{Gituliar:2017vzm}.
However, let us use this sector as a warm-up example and instead apply the method described in section~\ref{sec:eps-form_general}.

The differential equation for the three integrals of sector $\sectorb{}$ on the maximal cut has the following form,
\begin{equation}
\frac{\diff}{\diff x}\begin{pmatrix}
	\integralprec_2\\[0.1cm]
	\integralprec_3\\[0.1cm]
	\integralprec_4
\end{pmatrix}=
	\left(
\begin{array}{ccc}
 \frac{x^4 (5 \varepsilon -2)+x^2 (30 \varepsilon -8)+5 \varepsilon -2}{x (x^4-1)} & \frac{4 x (2 \varepsilon +1)}{(x^4-1) (4 \varepsilon -1)} & \frac{(x^2-1) \varepsilon }{x
   (x^2+1) (4 \varepsilon -1)} \\[0.2cm]
 \frac{4 (x^2-1) \varepsilon  (4 \varepsilon -1) (5 \varepsilon -1)}{x (x^2+1) (2 \varepsilon +1)} & \frac{4 (x^2-1) \varepsilon }{x^3+x} & -\frac{4 (x^2-1) \varepsilon
   ^2}{x (x^2+1) (2 \varepsilon +1)} \\[0.2cm]
 -\frac{(x^2-1) (4 \varepsilon -1) (5 \varepsilon -1)}{x^3+x} & \frac{4 x (2 \varepsilon +1)}{x^4-1} & \frac{x^4 (\varepsilon +1)+x^2 (6 \varepsilon +2)+\varepsilon +1}{x-x^5}
\end{array}
\right)\begin{pmatrix}
	\integralprec_2\\[0.1cm]
	\integralprec_3\\[0.1cm]
	\integralprec_4
\end{pmatrix}.
\end{equation}
To obtain a transformation of this block to an $\eps$-factorized form
we first make a choice for a seed integral $\integralseed^{(\sectorb{})}$.
As the sector is expected to be polylogarithmic, we would like to not only obtain an $\eps$-factorized form, but also such a form with entries being rational (and ideally with only simple poles).
A good choice turns out to be%
\footnote{Several other choices such as $\integralprec_{1010111111100000000000}$ also lead to an $\eps$-factorized form, although with a $\log$ dependence in the differential equation.}
\begin{equation}
	\integralseed^{(\sectorb{})}\equiv \integralprec_4= \integralprec_{1010112111100000000000}\,.
\end{equation}
Its leading singularity (LS) in $D=4$ (see refs.~\cite{Frellesvig:2024zph,Frellesvig:2024ymq} for details) takes a simple form given by 
\begin{equation}
\label{eq: LS_seed_b}
\text{LS}\left( \integralseed^{(\sectorb{})} \right) = \frac{x}{x^2-1}.
\end{equation}
In general dimensions, the integral has an associated Picard--Fuchs operator
\begin{equation}
\label{eq:PF_seed_sectorb}
\begin{aligned}
	\mathcal{L}_{\integralseed^{(\sectorb{})}}&=
	(x^2-1)^2 x^3 \frac{\diff^3}{\diff x^3}\\
	&\hphantom{{}={}}-2 (x^2-1) x^2 (4 x^2 \varepsilon -4 x^2+4 \varepsilon -1) \frac{\diff^2}{\diff x^2}\\
	&\hphantom{{}={}}+2 x (8 x^4 \varepsilon ^2-16 x^4 \varepsilon +7 x^4-56 x^2 \varepsilon ^2-8 x^2 \varepsilon -x^2+8 \varepsilon ^2-8
   \varepsilon +2) \frac{\diff}{\diff x}\\
   &\hphantom{{}={}}+4 (x^2-1) (x^2+1) (2 \varepsilon -1)^2,
\end{aligned}
\end{equation}
which in $D=4$ factorizes into a product of three different first-order operators 
\begin{equation}
	\mathcal{L}_{\integralseed^{(\sectorb{})}} \Big|_{\varepsilon=0} = \mathcal{L}_1 \cdot \mathcal{L}_1 \cdot \mathcal{L}_1\,.
\end{equation}
Since the coefficient $C_3(x,\eps)=(x^2-1)^2 x^3$ is independent of $\eps$, we can make an ansatz of the form
\begin{equation}
	J\frac{\diff}{\diff x}{\integralFinal}^{(\sectorb{})}=\eps  \left( \begin{array}{ccc}
	 F_{11} & F_{12} & 0 \\
	 F_{21} & F_{22} & F_{23} \\
	 F_{31} & F_{32} & F_{33} \\
 \end{array} \right){\integralFinal^{(\sectorb{})}}.
\end{equation}
Note that we are not imposing self-duality, as this turns out to be incompatible with the chosen seed integral; we will instead rotate to a self-dual form further below.

Following the notation of eq.~\eqref{PF_v1}, the operator $\mathcal{L}_{{\integralFinal}^{(\sectorb{})}_1}$ has coefficients
	\begin{align}
	\label{eq:sec_sectorbp_first}
		p_{3,0}&=J^3 F_{12}^2 F_{23}\,,\\
		p_{2,1}&=-J^2 F_{12}^2 F_{23} (F_{11}+F_{22}+F_{33})\,,\\
		p_{2,0}&=-J^2 F_{12} \left(F_{23} (2 J F_{12}'-3 F_{12} J')+J F_{12} F_{23}'\right)\,,\\
		p_{1,2}&=J F_{12}^2 F_{23} \left(F_{11} (F_{22}+F_{33})-F_{12} F_{21}+F_{22} F_{33}-F_{23} F_{32}\right)\,,\\
		p_{1,1}&=J F_{12} (F_{12} \left(J \left(F_{11}+F_{22}\right) F_{23}'-F_{23} \left(J' \left(F_{11}+F_{22}+F_{33}\right)+J \left(2 F_{11}'+F_{22}'\right)\right)\right)\nonumber\\*
		&\hphantom{{}={}}+J F_{23} F_{12}' \left(2F_{11}+F_{22}+F_{33}\right))\,,\\
		p_{1,0}&=J (J F_{12} F_{23}' \left(J F_{12}'-F_{12} J'\right)\nonumber\\*
		&\hphantom{{}={}}+F_{23} \left(J^2 \left(2 F_{12}'{}^2-F_{12} F_{12}''\right)+F_{12}^2 J'^2+J F_{12} \left(F_{12} J''-3 J' F_{12}'\right)\right))\,,\\
   		p_{0,3}&=F_{12}^2 F_{23} \left(F_{11} \left(F_{23} F_{32}-F_{22} F_{33}\right)+F_{12} \left(F_{21} F_{33}-F_{23} F_{31}\right)\right),\\
   		p_{0,2}&=J F_{12} (F_{12} \left(F_{23} \left(F_{11} F_{22}'+F_{33} F_{11}'\right)+F_{22} \left(F_{23} F_{11}'-F_{11} F_{23}'\right)\right)\nonumber\\*
   		&\hphantom{{}={}}-F_{11} F_{23} F_{12}' \left(F_{22}+F_{33}\right)+F_{12}^2
   \left(F_{21} F_{23}'-F_{23} F_{21}'\right))\,,\\
   		p_{0,1}&=J (-\left(F_{12}^2 \left(F_{23} J' F_{11}'+J \left(F_{23} F_{11}''-F_{11}' F_{23}'\right)\right)\right)\nonumber\\*
   		&\hphantom{{}={}}+F_{12} \left(F_{23} \left(F_{11} J' F_{12}'+2 J F_{11}' F_{12}'+J F_{11} F_{12}''\right)-J
   F_{11} F_{12}' F_{23}'\right)\nonumber\\*
   &\hphantom{{}={}}-2 J F_{11} F_{23} \left(F_{12}'\right){}^2)\,,\\
   		p_{0,0}&=0\,.   		
	\label{eq:sec_sectorbp_last}
	\end{align}
Note that the property ${p}_{0,0}=0$ was already observed as a requirement for ${\integralFinal}^{(\sectorb{})}_1$ to satisfy the uniform transcendentality property for polylogarithmic integrals in refs.~\cite{Hoschele:2014qsa,Dlapa:2020cwj,Dlapa:2022wdu}.

We now require that
\begin{equation}
\label{eq:sectorb_matching_integral}
	\integralFinal_1^{(\sectorb{})}=\frac{1}{\omega(x)}\integralseed^{(\sectorb{})}\,.
\end{equation}
The Picard--Fuchs operator of the right-hand side can easily be obtained from the one of $\integralseed^{(\sectorb{})}$ in eq.~\eqref{eq:PF_seed_sectorb}.
Following the notation of eq.~\eqref{PF_v2}, we find
	\begin{align}
	\label{eq:sec_sectorbptilde_first}
		\tilde{p}_{3,0}&=x^3 (x^2-1)^2 \omega\,,\\
		\tilde{p}_{2,1}&=-8 x^2 (x^4-1) \omega\,,\\
		\tilde{p}_{2,0}&=x^2 (x^2-1) \left(3 x (x^2-1) \omega '+(8 x^2+2) \omega\right),\\
		\tilde{p}_{1,2}&=16 x (x^4-7 x^2+1) \omega\,,\\
		\tilde{p}_{1,1}&=-16 x \left(x (x^4-1) \omega '+(2 x^4+x^2+1) \omega\right),\\
		\tilde{p}_{1,0}&=x \left(x (x^2-1) \left(3 x (x^2-1) \omega ''+4 (4 x^2+1) \omega '\right)+2 (7 x^4-x^2+2) \omega\right),\\
		\tilde{p}_{0,3}&=0\,,\\
		\tilde{p}_{0,2}&=16 \left((x^4-1) \omega+x (x^4-7 x^2+1) \omega '\right)\,,\\
		\tilde{p}_{0,1}&=-16 (x^4-1) \omega-8 x \left(x (x^4-1) \omega ''+2 (2 x^4+x^2+1) \omega '\right),\\
		\tilde{p}_{0,0}&=4 (x^4-1) \omega+x (x (x^2-1) \left(x (x^2-1) \omega'''+(8 x^2+2) \omega ''\right)\nonumber\\*
		&\hphantom{{}={}}+2 (7 x^4-x^2+2) \omega ')\,,
		\label{eq:sec_sectorbptilde_last}
	\end{align}
The operators of both sides of eq.~\eqref{eq:sectorb_matching_integral} only have to match up to a function $c(x)$, such that we require
\bq
	\label{eq:sectorb_matching_condition}
 \tilde{p}_{k,j}\left(x\right) & = & c\left(x\right) \cdot p_{k,j}\left(x\right),\quad 0\le k\le 3,\  0\le j\le 3-k.
\eq
Through this identification, we can derive constraints for the free parameters of the ansatz, in this case
\begin{equation}
	F_{11},\ F_{12},\ F_{21},\ F_{22},\ F_{23},\ F_{31},\ F_{32},\ F_{33},\ J,\ \omega,\ c.
\end{equation}
In solving these constraints, we can proceed systematically from largest to smallest $k$, for each progressing from largest to smallest $j$.
This corresponds to the order of coefficients in eqs.~\eqref{eq:sec_sectorbp_first}--\eqref{eq:sec_sectorbp_last} and \eqref{eq:sec_sectorbptilde_first}--\eqref{eq:sec_sectorbptilde_last}.
In each case, we first eliminate as many parameters as possible using the previously obtained solutions, and then solving for one of the remaining free parameters or their derivatives that appear linearly in the constraint.
If a desired parameter appears both directly and via its derivative, we have to solve for the highest derivative appearing in the constraint.
For the case at hand, we proceed as detailed on in table~\ref{tab: strategy subsector}.
\begin{table}[b]\centering$\begin{NiceArray}{c|cccc}[columns-width = 2em]
\diagbox{k}{j} & 0 & 1 & 2 & 3 \\ \hline
3 & c &  &  &  \\
2 & F_{12}' & F_{11} &  &  \\
1 & F_{23}'' & F_{22}' & F_{21} &  \\
0 & \omega''' & F_{33}'' & F_{32}' & F_{31}
\end{NiceArray}$
\caption{Strategy for which free parameter to solve for in the matching equation~\eqref{eq:sectorb_matching_condition}, for a given $k$ and $j$.}
\label{tab: strategy subsector}
\end{table}
Note that for a given coefficient $p_{k,j}$ we obtain a constraint for the $(3-k-j)$th derivative of a given parameter. 
Furthermore, labeling $c\equiv F_{01}$ and $\omega\equiv F_{34}$ we find that the constraint for coefficient $p_{k,j}$ can be conveniently solved for $F_{3-k,4-k-j}$.
The Jacobian $J$ is not fixed by the ansatz and can be chosen freely.%
\footnote{We can use $J$ to normalize one of the entries of our ansatz. In refs.~\cite{Pogel:2022ken,Pogel:2022vat}, this freedom was used to set $F_{12}=1$.}
In the following, we thus assume $J=1$.
We list some of the simpler differential constraints for the free parameters:
\begin{align}
	c&= \frac{x^3 (x^2-1)^2 \omega }{F_{12}^2 F_{23}}\,,\\
	F_{12}'&= \frac{1}{2} F_{12} \left(-\frac{F_{23}'}{F_{23}}+\frac{8 x^2+2}{x-x^3}-\frac{3 \omega '}{\omega }\right),\\
		F_{11}&=\frac{\left(x-x^3\right) F_{22}+\left(x-x^3\right) F_{33}+8 (x^2+1)}{x (x^2-1)}\,,\\
		\omega'''&= \frac{2 x \left(x \left(-4 x^4+3 x^2+1\right) \omega ''+\left(-7 x^4+x^2-2\right) \omega '\right)-4 (x^4-1) \omega }{x^3 (x^2-1)^2}\,,\\
		F_{23}''&=  F_{23}\left(\frac{ \frac{2 \left(4 x^2+1\right) \omega '}{x (x^2-1)}+3 \omega ''}{\omega }+\frac{2 \left(2 x^4-3 x^2+2\right)}{x^2 (x^2-1)^2}-\frac{3 \omega '^2}{2\omega ^2}\right)+\frac{3}{2}\frac{
  F_{23}'{}^2}{F_{23}}\,.\\
   	&\vdots \nonumber
	\end{align}
The condition $c\cdot p_{0,0}=0=\tilde{p}_{0,0}$ translates directly to the requirement that $\omega$ is annihilated by the Picard--Fuchs operator of the seed integral in the limit $\eps\to 0$.
This corresponds to the common practice of normalizing integrals in an $\eps$-factorized basis by their leading singularity.
In the case of elliptic and Calabi--Yau integrals, this is equivalent to normalizing the integral by a period of the associated geometry.

We can now go ahead and solve the constraints we obtained for the parameters $F_{ij}$, $c$, and $\omega$.
While at first glance the constraints appear coupled, they do possess a triangular structure, and can be solved individually given the right order.
In the given case, we can solve for the free parameters of the ansatz in the following order:
\begin{equation*}
	\omega\to F_{23}\to F_{12}\to c\to F_{33}\to F_{22} \to F_{11} \to F_{32} \to F_{21}\to F_{31}\,.
\end{equation*}
Considering table~\ref{tab: strategy subsector}, this order corresponds to solving the constraints from lowest to highest $k$, from lowest to highest $j$.

As we expect sector $\sectorb{}$ to be polylogarithmic based on the leading singularity analysis, we expect the constraints to be solvable in terms of rational functions of $x$.
Indeed, choosing appropriate boundary conditions we find the following solutions:
\begin{equation}
\begin{gathered}
	\omega = \frac{x}{x^2-1},\quad 
	F_{23}= \frac{4}{x},\quad 
	F_{12}= \frac{4 \sqrt{5}}{x^2-1},\quad 
	c = \frac{1}{320} x^5 (x^2-1)^3,\quad
	F_{33}= 0,\\ 
	F_{22}= \frac{8 (x^2+1)}{x(x^2-1)},\quad 
   F_{11}= 0,\quad 
   F_{32}= -\frac{4}{x},\quad 
   F_{21}= \frac{4 \sqrt{5}}{x^2-1},\quad 
   F_{31}= 0,\quad 
   J= 1,
\end{gathered}
\end{equation}
where we note that $\omega$ precisely matches the leading singularity calculated in eq.~\eqref{eq: LS_seed_b}, as expected. The $\eps$-factorized differential equation is then
\begin{equation}
	\frac{\diff}{\diff x}{\integralFinal}^{(\sectorb{})}=\varepsilon\left(
\begin{array}{ccc}
 0 & \hspace{0.2cm} \frac{4 \sqrt{5}}{x^2-1} & \hspace{0.2cm} 0 \\[0.1cm]
 \frac{4 \sqrt{5}}{x^2-1} & \hspace{0.2cm} -\frac{8}{x}+\frac{8}{x+1}+\frac{8}{x-1} & \hspace{0.2cm} \frac{4}{x} \\[0.1cm]
 0 & \hspace{0.2cm} -\frac{4}{x} & \hspace{0.2cm} 0 \\
\end{array}
\right){\integralFinal}^{(\sectorb{})},
\end{equation}
where ${\mathcal{J}}^{(\sectorb{})}=U^{(\sectorb{})} (\integralprec_2,\integralprec_3,\integralprec_4)^T$, with
\begin{equation}
	U^{(\sectorb{})}=\left(
\begin{array}{ccc}
 0 & 0 & x-\frac{1}{x} \\[0.3cm]
 -\frac{(x^2-1)^3 (4 \varepsilon -1) (5 \varepsilon -1)}{4 \sqrt{5} x^2 (x^2+1) \varepsilon } & \frac{(x^2-1) (2 \varepsilon +1)}{\sqrt{5} (x^2+1) \varepsilon } &
 -\frac{(x^2-1) (x^4+6 x^2+1)}{4 \sqrt{5} x^2 (x^2+1)} \\[0.3cm]
 \frac{(x^2-1)^2 (4 \varepsilon -1) (5 \varepsilon -1)}{4 \sqrt{5} x^2 \varepsilon } & -\frac{(2 \varepsilon +1) (10 \varepsilon -1)}{4 \sqrt{5} \varepsilon ^2} & \frac{(x^2-1)^2}{4 \sqrt{5}
   x^2} \\
\end{array}
\right).
\end{equation}

While we were not able to find a suitable seed integral to impose self-duality from the start, we can see that the differential equation is suggestive of such a form, including a possible Galois symmetry \cite{Pogel:2024sdi}.
By ansatzing a constant rotation of our master integrals, we can bring the differential equation into a manifestly self-dual and Galois symmetric form.
For $\integralFinalReally^{(\sectorb{})}=\tilde{U}^{(\sectorb{})}\integralFinal^{(\sectorb{})}$ with 
\begin{equation}
	\tilde{U}^{(\sectorb{})}=\left(
\begin{array}{ccc}
 1 & i & 0 \\
 0 & 0 & i \sqrt{2} \\
 1 & -i & 0 \\
\end{array}
\right),
\end{equation}
we find
\begin{equation}	\diff\integralFinalReally^{(\sectorb{})}=\varepsilon A^{(\sectorb{})}\integralFinalReally^{(\sectorb{})}=\varepsilon\scalebox{1}{$\left(
\begin{array}{ccc}
 4 f_2 & 2 \sqrt{2} f_1 & -2(2 f_2- \sqrt{-5} f_3) \\
 -2 \sqrt{2} f_1 & 0 & 2 \sqrt{2} f_1 \\
 -2(2 f_2+ \sqrt{-5} f_3) & -2 \sqrt{2} f_1 & 4 f_2 \\
\end{array}
\right)$}\integralFinalReally^{(\sectorb{})},
\end{equation}
where
\begin{equation}
	f_1=\diff\log\left(x\right),\quad f_2=\diff\log\left(\frac{1-x^2}{x}\right),\quad 
	f_3=\diff\log\left(\frac{1-x}{1+x}\right).
\end{equation}
The differential equation $A^{(\sectorb{})}$ is self-dual, i.e.~symmetric along its anti-diagonal.
Further defining the Galois action
\begin{equation}
\label{eq: Galois action}
	\rho_{-5,2}: \begin{array}[t]{rl}
		\sqrt{-5}&\mapsto -\sqrt{-5}\\
		\sqrt{2} &\mapsto -\sqrt{2}\,,
	\end{array}
\end{equation}
it transforms as
\begin{equation}
	\rho_{-5,2}\left[\tilde{A}^{(\sectorb{})}\right]=(\tilde{A}^{(\sectorb{})})^T\,.
\end{equation}

Finally, we complete the subsectors by bringing the off-diagonal $1\times 3$ block linking sectors $\sectorb{}$ and $\sectorc{}$ into $\eps$-factorized form.
After the rotations of sector $\sectorb{}$, the entries in the off-diagonal are already $\eps$-factorized, however with double poles. We reduce them to single poles using \texttt{Fuchsia} \cite{Gituliar:2017vzm}, and perform an additional rotation to eliminate the dependence on $\integralFinal^{(\sectorb{})}_1$ and $\integralFinal^{(\sectorb{})}_3$.
With 
\begin{equation}
\begin{aligned}
	\integralFinalReally^{(\text{sub})}=\{\integralFinalReally_1^{(\sectora{})},
		\integralFinalReally_1^{(\sectorb{})},
		\integralFinalReally_2^{(\sectorb{})},
		\integralFinalReally_3^{(\sectorb{})},
		\integralFinalReally_1^{(\sectorc{})}\}\,,
\end{aligned}
\end{equation}
we thus find
\begin{equation}
\frac{\diff}{\diff x}
	\integralFinalReally^{(\text{sub})}=\varepsilon
	\left(
\begin{array}{ccccc}
 0 & 0 & 0 & 0 & 0 \\
 0 & 4 f_2   & 2 \sqrt{2} f_1   & -2(2 f_2  - \sqrt{-5} f_3 )  & 0 \\
 0 & -2 \sqrt{2} f_1   & 0 & 2 \sqrt{2} f_1   & 0 \\
 0 & -2 \left(2 f_2+ \sqrt{-5} f_3\right)   & -2 \sqrt{2} f_1   & 4 f_2   & 0 \\
 0 & 0 & f_2   & 0 & -4 f_2   \\
\end{array}
\right)\integralFinalReally^{(\text{sub})}\,,
\end{equation}
for
\begin{equation}
	\integralFinalReally^{(\text{sub})}=\underbrace{ c_2
	\scalebox{0.8}{$
	\left(
\begin{array}{ccccc}
 \frac{c_1}{c_2} & 0 & 0 & 0 & 0 \\[0.2cm]
 0 & \frac{ (x^2-1)^3 (4 \varepsilon -1) (5 \varepsilon -1)}{4 \sqrt{-5} x^2 (x^2+1) \varepsilon } & -\frac{ (x^2-1) (2 \varepsilon +1)}{\sqrt{-5} (x^2+1) \varepsilon } &
   \frac{ (x^2-1) (x^4+6 x^2+1)}{4 \sqrt{-5} (x^2+1) x^2}+x-\frac{1}{x} & 0 \\[0.2cm]
 0 & -\frac{ (x^2-1)^2 (4 \varepsilon -1) (5 \varepsilon -1)}{2 \sqrt{2}\sqrt{-5} x^2 \varepsilon } & \frac{ (2 \varepsilon +1) (10 \varepsilon -1)}{2 \sqrt{2}\sqrt{-5} \varepsilon ^2} & -\frac{ (x^2-1)^2}{2
   \sqrt{2}\sqrt{-5} x^2} & 0 \\[0.2cm]
 0 & -\frac{ (x^2-1)^3 (4 \varepsilon -1) (5 \varepsilon -1)}{4 \sqrt{-5} x^2 (x^2+1) \varepsilon } & \frac{ (x^2-1) (2 \varepsilon +1)}{\sqrt{-5} (x^2+1) \varepsilon } &
   -\frac{ (x^2-1) (x^4+6 x^2+1)}{4 \sqrt{-5} (x^2+1) x^2}+x-\frac{1}{x} & 0 \\[0.2cm]
 0 & -\frac{(x^2-1)^2 (4 \varepsilon -1) (5 \varepsilon -1)}{16 \sqrt{2}\sqrt{-5} x^2 \varepsilon } & \frac{ (2 \varepsilon +1) (26 \varepsilon -3)}{32\sqrt{2} \sqrt{-5} \varepsilon ^2} & -\frac{
   (x^2-1)^2}{16 \sqrt{2}\sqrt{-5} x^2} & -\frac{ (1-6 \varepsilon ) (14 \varepsilon -1)}{16 \sqrt{2}\sqrt{-5} \varepsilon } \\
\end{array}
\right)$}}_{\tilde{U}^{(\text{sub})}}
	\begin{pmatrix}
		\integralprec_1\\[0.1cm]
		\integralprec_2\\[0.1cm]
		\integralprec_3\\[0.1cm]
		\integralprec_4\\[0.1cm]
		\integralprec_5
	\end{pmatrix},
\end{equation}
where we further introduced the normalizations
\begin{equation}
	c_1=\frac{(1-6 \varepsilon )^2}{2 \varepsilon +1},\quad c_2=\frac{\varepsilon ^5}{(4 \varepsilon +1)^2 (6 \varepsilon +1)^2 (14 \varepsilon -1)}
\end{equation}
for later convenience.
\subsection{Topsector}
Now that we have demonstrated our procedure on the simple example of the subsector integrals, let us apply it to the topsector, sector $\sectord{}$.
As the procedure is for the most part identical, we will only highlight the novel features appearing for this sector.

We pick as our seed integral 
\begin{equation}
	\integralseed^{(\sectord{})}=\integralprec_6=\integralprec_{111 111 111 11 000 000 000 00}\,,
\end{equation}
which in $D$ dimensions is annihilated by the fifth-order Picard--Fuchs operator~\eqref{eq:PF_L5}, and in 4 dimensions by the fourth-order operator~\eqref{eq: PF_L4_4d}.
In deriving an $\eps$-factorized differential equation, the fifth-order operator is the relevant one, and we need to make an ansatz for a differential equation that is compatible with the operator.
As discussed in section \ref{sec:analysis_CY}, the main novel feature which we would like to incorporate in our ansatz is an apparent singularity, depending quadratically on $\eps$.
For a fifth-order operator, such an ansatz was already suggested in section \ref{sec:eps-form_general}, see eq.~\eqref{eq:ansatz_section3}. 
We assume that the final master integrals $\integralFinal^{(\sectord{})}=(\integralFinal^{(\sectord{})}_1,\dots,\integralFinal^{(\sectord{})}_5)^T$ satisfy the manifestly self-dual differential equation%
\footnote{Note that the functions $F_{ij}$, $\omega$, and $J$ are different for each sector.}
\begin{equation}
\label{eq:sector_d_ansatz}
	J\frac{\diff}{\diff x}\integralFinal^{(\sectord{})}=\eps   \left( \begin{array}{ccccc}
 F_{11} & F_{12} & 0 & 0 & 0 \\
 F_{21} & F_{22} & F_{23} & F_{24} & 0 \\
 F_{31} & F_{32} & F_{33} & F_{23} & 0 \\
 F_{41} & F_{42} & F_{32} & F_{22} & F_{12} \\
 F_{51} & F_{41} & F_{31} & F_{21} & F_{11} \\
 \end{array} \right)\integralFinal^{(\sectord{})}
\end{equation}
and assume
\begin{equation}
\label{eq:sectorsectorb_matching_integral}
	\integralFinal_1^{(\sectord{})}=\frac{\varepsilon}{\omega(x)}\integralseed^{(\sectord{})}\,.
\end{equation}
Matching the Picard--Fuchs operators, see eqs.~\eqref{PF_v1} and~\eqref{PF_v2}, we find $33$ constraints of the form
\bq
	\label{eq:sectord_matching_condition}
 \tilde{p}_{k,j}\left(x\right) & = & c\left(x\right) \cdot p_{k,j}\left(x\right),\quad 0\le k\le 5,\  0\le j\le 7-k.
\eq
We solve these constraints again in the order from greatest to smallest $k$, and from greatest to smallest $j$,
for the parameters shown in table \ref{tab: solve_order_sectord}.
\begin{table}\centering 
$
\begin{NiceArray}{c|cccccccc}[columns-width = 2em]
\diagbox{k}{j} & 0 & 1 & 2 & 3 & 4& 5& 6& 7    \\ \hline
5 & F_{21} & F_{11} & c & &  &  &  & \\
4 & 0 & 0 & J'' & F_{22}'&  &  &  & \\
3 & F_{33} & F_{24}' & F_{23}''' & F_{33}' & F_{31} &  &  & \\
2 & 0 & 0 & 0 & F_{32}'' & 0 & F_{41} &  &  \\
1 & 0 & \omega'''' & 0 & 0 & F_{42}''  & 0 & F_{51} &  \\
0 & 0 & 0 & 0 & 0 & 0 & 0 & 0 & F_{24}
\end{NiceArray}$
\caption{Strategy for which free parameter to solve for in the matching equation~\eqref{eq:sectord_matching_condition}, for a given $k$ and $j$.}
\label{tab: solve_order_sectord}
\end{table}
The parameter $F_{12}$ is unconstrained and can be set to $1$. 

Note that the order of parameters for which we solve for is slightly less systematic than in the subsector example.
This has three main reasons:
First, through the manifestly self-dual ansatz we already assumed that certain parameters are equal.
Thus certain constraints have to be solved for different parameters, or are already satisfied; the latter cases are denoted by zeroes in table~\ref{tab: solve_order_sectord}.
Second, the elimination of dependencies using previously solved parameters can lead to large intermediate expressions.
While these typically  simplify greatly, this simplification can be computationally expensive. 
It may end up being simpler to give up any systematic order and solve the constraints for parameters which later on cause less bloat in expressions.
This is for example the reason that we chose to solve constraints for $J''$ instead of $F_{12}$.\footnote{Note, however, that this choice is only a question of the normalization of the differential equation.} 
Third, it can be the case that several constraints lead to the same solutions, with one of them being easier to simplify than the others.
An example of this are the constraints $k=1,j=1$ and $k=1,j=0$ in the present case.
While both lead to exactly the same solution for $\omega''''$, this solution is significantly easier to obtain from the latter constraint.
We again show some of the easier constraints and provide the full list in computer readable form in an ancillary file.
\begin{align}
		\omega''''&=-\frac{2-16 x^2-10 x^4}{x(1-x^4)}\omega'''-\frac{1-28 x^2+46 x^4+68 x^6+25 x^8}{x^2 (1-x^4)^2} \omega ''\nonumber\\*
   &\hphantom{{}={}}+\frac{1+11 x^2-54 x^4+22 x^6+37 x^8+15 x^{10}}{x^3 (1-x^2)^3
   (1+x^2)^2}\omega '\\*
   &\hphantom{{}={}}-\frac{1+3 x^2+20 x^4+3 x^6+x^8}{x^4 (1-x^4)^2}\omega,\nonumber\\
   F_{33}&= \frac{F_{23}}{F_{24}}+\frac{40 x J}{x^4-1},\\
   F_{11}&= \frac{2 F_{24} \left((x-x^5) F_{22}+2 J (x^4-8 x^2+1)\right)+(x-x^5)
   F_{23}^2}{2 x (x^4-1) F_{24}},\\
   F_{24}'&= F_{24} \left(\frac{3 J'}{J}-\frac{-5 x^4-8 x^2+1}{x-x^5}-\frac{2 \omega '}{\omega }\right).\\
   &\vdots \nonumber
\end{align}

Already at this stage we can observe that geometric data from the Calabi--Yau three-variety makes a prominent appearance.
Inspecting the solution for $\omega''''$ and comparing to the operator $\mathcal{L}_4$ in eq.~\eqref{eq: PF_L4_4d}, we see that the manifestly self-dual ansatz forces us to choose a period of the Calabi--Yau three-variety as normalization of the first integral.
For a generic ansatz this is not the case, and one could also choose the fifth solution of the fifth order operator in the limit $\eps\to 0$ as normalization.
The constraint on $F_{24}'$ turns out to be precisely the differential equation for the $Y$-invariant $Y_2$ of eq.~\eqref{eq:Y2}, which is just the Yukawa coupling of the underlying Calabi--Yau three-variety.
Finally, the constraint on $J''$ requires that $J$ is proportional to the Jacobian obtained when changing derivatives in $x$ to logarithmic derivatives in the mirror map $q$ of eq.~\eqref{eq:q-coord}, i.e.~$J \frac{\diff}{\diff x}=\theta_q$ or $J=\frac{\diff x}{\diff \log(q)}$.
Similar observations were made in refs.~\cite{Pogel:2022ken,Pogel:2022vat,Duhr:2022dxb} in the case of banana and ice cone integrals.

As in the subsector, the constraints we obtain appear to be highly coupled.
However, just as before, it is possible to choose an ordering of parameters such that they can be solved independently.
One such ordering is
\begin{equation}
	\omega\to J\to F_{23}\to \{F_{24},c,F_{33}\} \to F_{32}\to \{F_{22},F_{42}\}\to \{F_{11},F_{21},F_{31},F_{41},F_{51}\}\,,
\end{equation}
where parameters in braces can be solved in any order.
As there is no known class of differential forms associated to the underlying Calabi--Yau geometry in which we could express the $F_{ij}$, we resort to deriving series expansions solutions for them.
Requiring that 
\begin{equation}
	\omega=\varpi_1=\Sigma_1\,,\quad F_{24}=Y_2\,,\quad J=\frac{\diff x}{\diff \log(q)}\,,
\end{equation}
with $\varpi_1$ being the holomorphic Frobenius basis element of eq.~\eqref{eq:Frobenius series},
naturally introduces the object $\sqrt{\tfrac{-5}{2}}$, made up to of the two roots with respect to which we observed a Galois symmetry in the subsector $\sectorb$.
By making appropriate boundary choices in solving for the parameters $F_{ij}$, we can limit the dependence on this root to the parameters $F_{23}$, $F_{31}$ and $F_{32}$, such that
\begin{align}
	P&=-\tfrac{1}{10}\sqrt{\tfrac{-5}{2}}\left(\frac{1}{x}+\frac{13 x}{2^3}+\frac{3441 x^3}{2^{11}}+\frac{25343 x^5}{2^{14}}+\frac{190544849 x^7}{2^{27}}+\mathcal{O}(x^9)\right),\\
	\omega &=x+\frac{9 x^3}{2^4}+\frac{1681 x^5}{2^{12}}+\frac{21609 x^7}{2^{16}}+\mathcal{O}(x^9)\,,\\
	J&=x-\frac{3 x^3}{2^2}-\frac{77 x^5}{2^{10}}-\frac{369 x^7}{2^{13}}+\mathcal{O}(x^9)\,,\\
	F_{11}&=-1+\frac{15 x^2}{2^2}+\frac{1005 x^4}{2^{10}}+\frac{5505 x^6}{2^{13}}+\frac{33070845 x^8}{2^{26}}+\mathcal{O}(x^{9})\,,\\
	F_{21}&=1+\frac{171 x^2}{2^2}-\frac{18809 x^4}{2^{10}}+\frac{37767 x^6}{2^{13}}+\frac{206196007 x^8}{2^{26}}+\mathcal{O}(x^9)\,,\\
	F_{22}&=\frac{1}{4}\left(1+\frac{705 x^2}{2^3}-\frac{76795 x^4}{2^{11}}+\frac{4575 x^6}{2^{14}}-\frac{242517275 x^8}{2^{27}}+\mathcal{O}(x^9)\right),\\
	F_{23}&= \sqrt{\tfrac{-5}{2}}\left(1+\frac{29 x^2}{2^3}-\frac{3495 x^4}{2^{11}}-\frac{1225 x^6}{2^{14}}-\frac{19494119 x^8}{2^{27}}+\mathcal{O}(x^9)\right),\\
	F_{24}&= 1+\frac{5 x^2}{2^3}-\frac{775 x^4}{2^{11}}-\frac{445 x^6}{2^{14}}-\frac{5110375 x^8}{2^{27}}+\mathcal{O}(x^9)\,,\\
	F_{31}&=-\sqrt{\tfrac{-5}{2}}\left(1+\frac{3 x^2}{2^2}-\frac{264601 x^4}{2^{10}}-\frac{988629 x^6}{2^{13}}-\frac{7257154393 x^8}{2^{26}}+\mathcal{O}(x^9)\right),\\
	F_{32}&=-\tfrac{1}{4}\sqrt{\tfrac{-5}{2}}\left(1+\frac{3961 x^2}{2^3}-\frac{321563 x^4}{2^{11}}+\frac{234179 x^6}{2^{14}}+\frac{77183589 x^8}{2^{27}}+\mathcal{O}(x^9)\right),\\
	F_{33}&=-\frac{5}{2}\left(1+\frac{181 x^2}{2^3}-\frac{12359 x^4}{2^{11}}+\frac{17707 x^6}{2^{14}}+\frac{54628505 x^8}{2^{27}}+\mathcal{O}(x^9)\right),\\
	F_{41}&=\frac{5}{4}\left(1-\frac{1189 x^2}{4}-\frac{1225145 x^4}{2^{10}}-\frac{4233457 x^6}{2^{13}}-\frac{29885486041 x^8}{2^{26}}+\mathcal{O}(x^9)\right),\\
		F_{42}&=-\frac{15}{2^4}+\frac{84429 x^2}{2^7}-\frac{8120415 x^4}{2^{15}}+\frac{2919659 x^6}{2^{18}}-\frac{11826290495 x^8}{2^{31}}+\mathcal{O}\left(x^9\right),\\
	F_{51}&=\frac{881 x^2}{2}+\frac{1734625 x^4}{2^9}+\frac{23147105 x^6}{2^{12}}+\frac{191659062625 x^8}{2^{25}}+\mathcal{O}(x^9)\,.
\end{align}
Modulo their overall algebraic or rational prefactors, these series solutions are $N$-integral with $N=2^4$.

The transformation $U^{(\sectord)}$ such that $\integralFinal^{(\sectord)}=U^{(\sectord)}\integralprec^{(\sectord)}$ is expressible in terms of the $F_{ij}$, $\omega$, and $J$.
With the solutions above, it has the following form%
\footnote{We provide full expressions in the ancillary file.}
\begin{equation}
	U^{(\sectord)}=
	\text{diag}\Bigl(1,1,\sqrt{\tfrac{-5}{2}},1,1\Bigr)
	\begin{pNiceMatrix}[columns-width = 1em]
		\frac{1}{\varpi_1} & 0 & 0 & 0 & 0 \\
		\bullet & \bullet & \bullet & \bullet & 0 \\
		\bullet & \bullet & \bullet & \bullet & \tfrac{1}{\varrho} \\
		\bullet & \bullet & \bullet & \bullet & \bullet\\
		\bullet & \bullet & \bullet & \bullet & \bullet\\
	\end{pNiceMatrix}
	\text{diag}\Bigl(1,\tfrac{2 \varepsilon +1}{6 \varepsilon +1},\tfrac{2 \varepsilon +1}{\varepsilon  (6 \varepsilon +1)},\tfrac{4 \varepsilon +1}{6 \varepsilon +1}, 6 \varepsilon -1\Bigr),
\end{equation}
where 
\begin{equation}
	\varrho=\frac{x}{1-x^2}
\end{equation}
is the leading singularity of the integral $\integralprec_{10}$,%
\footnote{Concretely, the Picard--Fuchs operator for $\integralprec_{10}$ factors in four dimensions, $\mathcal{L}_5\big|_{\varepsilon=0} =  \mathcal{L}_4\cdot\mathcal{L}_1$, and $\varrho$ is the solution to the first-order factor $\mathcal{L}_1$.
}
 and $\bullet$ represents other non-zero entries.
In addition, we find 
\begin{equation}
\begin{aligned}
	U^{(\sectord)}_{3j}&=\sqrt{\tfrac{-5}{2}}f(x) U^{(\sectord)}_{2j},\quad j=2,3,4\,,\\
	U^{(\sectord)}_{31}&=\sqrt{\tfrac{-5}{2}}\left(f(x) U^{(\sectord)}_{21}+g(x,\eps)U^{(\sectord)}_{21}\right),
\end{aligned}
\end{equation} 
with
\begin{equation}
\begin{aligned}
	f(x) &= 3-\frac{3 x^2}{2^3}-\frac{283 x^4}{2^{11}}-\frac{2567 x^6}{2^{15}}-\frac{7084667 x^8}{2^{27}}+\mathcal{O}\left(x^9\right),\\
	g(x,\eps) &= \frac{1}{\eps}\Big((1-4 \varepsilon) - \frac{1}{2^4} x^2 (116 \varepsilon + 3) - \frac{x^4 (14852 \varepsilon + 231)}{2^{12}}\\
	&\hspace{2em} - \frac{x^6 (158708 \varepsilon + 1819)}{2^{16}} - \frac{x^8 (488403524 \varepsilon
   + 4487967)}{2^{28}}+\mathcal{O}(x^9)\Big).
\end{aligned}
\end{equation}
The functions $f(x)$ and $g(x,\eps)$ again have an $N$-integral series expansion with $N=2^4$.
Thus, the third master integral $\integralFinal_3^{(\sectord)}$ can be thought of to take the form
\begin{equation}
\label{eq:Jd_3}
	\integralFinal_3^{(\sectord)}=\sqrt{\tfrac{-5}{2}}\left(\frac{6\eps-1}{\varrho}\integralprec_{10}+f(x)\, \integralFinal_2^{(\sectord)}+g(x,\eps) \integralFinal_1^{(\sectord)}\right).
\end{equation}
Based on this identification, the form of the differential equation of eq.~\eqref{eq:sector_d_ansatz}, as well as $F_{12}=1$, our extended ansatzing method directly leads to a basis of the form,
\begin{equation}
\begin{aligned}
	\integralFinal_{1}^{(\sectord)}&=\frac{\eps^3}{\omega}\integralseed^{(\sectord)}\,,\\
	\integralFinal_{2}^{(\sectord)}&=\frac{J}{\eps}\frac{\diff \integralFinal_{1}^{(\sectord)}}{\diff x}-F_{11}\integralFinal_{1}^{(\sectord)}\,,\\
	\integralFinal_3^{(\sectord)}&=C_1\,\mathcal{I}_{\text{extra}}+C_2\,\integralFinal_{1}^{(\sectord)}+C_3\, \integralFinal_{2}^{(\sectord)}\,,\\
	\integralFinal_{4}^{(\sectord)}&=\frac{1}{F_{24}}\left(\frac{J}{\eps}\frac{\diff \integralFinal_{2}^{(\sectord)}}{\diff x}-F_{21}\integralFinal_{1}^{(\sectord)}-F_{22}\integralFinal_{2}^{(\sectord)}-F_{23}\integralFinal_{3}^{(\sectord)}\right),\\
	\integralFinal_{5}^{(\sectord)}&=\frac{J}{\eps}\frac{\diff \integralFinal_{4}^{(\sectord)}}{\diff x}-F_{41}\integralFinal_{1}^{(\sectord)}-F_{42}\integralFinal_{2}^{(\sectord)}-F_{43}\integralFinal_{3}^{(\sectord)}-F_{44}\integralFinal_{4}^{(\sectord)}\,,
\end{aligned}
\end{equation}
for some functions $C_1$, $C_2$, $C_3$ which can easily be read off from eq.~\eqref{eq:Jd_3}. 
Considering that $\omega=\varpi_1$, the Jacobian $J=\frac{\diff x}{\diff \log(q)}$, and $F_{24}=Y_2$ can be related to Calabi--Yau data, this basis resembles a standard derivative basis, as was for example employed for other Calabi--Yau three-variety integrals such as the banana integral in ref.~\cite{Pogel:2022ken}, with an additional integral depending on $\mathcal{I}_{\text{extra}}$---in our case $\integralprec_{10}$---inserted as third master integral.
Similar insertions are commonly known to appear also in elliptic Feynman integrals, where an elliptic sector is extended by integrals that decouple for $\eps\to 0$,
leading to the appearance of elliptic integrals of the third kind, see for example refs.~\cite{Remiddi:2013joa,Tancredi:2015pta,Duhr:2022dxb,Gorges:2023zgv}.
Considering the differential equation of the precanoncial basis $\integralprec$, the additional integral $\mathcal{I}_{10}$ can be observed to decouple from $\integralprec_{5}$--$\integralprec_{9}$ in this limit.
It is the only master integral in sector $\sectord$ with an ISP in the numerator, which corresponds to a pole at infinity.
The normalization $\varrho$ that we find for $\integralprec_{10}$ in our $\eps$-factorized basis corresponds to the leading singularity obtained by taking the residue at this pole at infinity.
Thus, the integral $\integralprec_{10}$ could be interpreted as a Calabi--Yau analogue of elliptic integrals of the third kind.

Finally, we are left with only the off-diagonal $5\times 5$ block highlighted in red in eq.~\eqref{eq:A_prec}, coupling the subsectors to the topsector.
After the rotation to
\begin{equation}
	\begin{pmatrix}
		\integralFinalReally^{(\text{sub})}\\
		\integralFinal^{(\sectord)}
	\end{pmatrix}=\left(\begin{NiceArray}{c|c}[]
	\tilde{U}^{(\text{sub})}& 0 \\ \hline
	0& U^{(\sectord)}
	\end{NiceArray}\right)\integralprec\,,
\end{equation}
all the entries of the differential equation become a Laurent series in $\eps$ and truncate at linear order $\eps^1$. 
The entries further have the following leading order behavior in $\eps$:
\begin{equation}
	\frac{\diff}{\diff x}\begin{pmatrix}
		\integralFinalReally^{(\text{sub})}\\
		\integralFinal^{(\sectord)}
	\end{pmatrix}=A^{(\text{diag})}_x(x,\eps)\begin{pmatrix}
		\integralFinalReally^{(\text{sub})}\\
		\integralFinal^{(\sectord)}
	\end{pmatrix}\sim
	\scalebox{0.7}{$
	\left(\begin{NiceArray}{ccccc|ccccc}[columns-width = 1.5em]
	0 & &&&&&&&&\\
	 &\eps & \eps &\eps & &&&&\\	
	 &\eps & 0 &\eps & &&&&\\	
	 &\eps & \eps &\eps & &&&&\\
	 & & \eps &  & \eps&&&&\\\hline
	\eps &\eps^{-2} &\eps^{-3} & \eps^{-2} &\eps^{-3} & \eps & \eps & & & \\
	\eps^0 &\eps^{-3} &\eps^{-4} & \eps^{-3} &\eps^{-4} & \eps & \eps & \eps & & \\
	\eps^0 &\eps^{-4} &\eps^{-4} & \eps^{-4} &\eps^{-4} & \eps & \eps & \eps &  \eps& \\
	\eps^{-1} &\eps^{-4} &\eps^{-5} & \eps^{-4} &\eps^{-5} & \eps & \eps & \eps & \eps & \eps\\
	\eps^{-2} &\eps^{-5} &\eps^{-6} & \eps^{-5} &\eps^{-6} & \eps & \eps & \eps & \eps & \eps \\
	\end{NiceArray}\right)$}
	\begin{pmatrix}
		\integralFinalReally^{(\text{sub})}\\
		\integralFinal^{(\sectord)}
	\end{pmatrix}\,.
\end{equation}
The columns in the off-diagonal block belonging to $\integralFinalReally_2^{(\text{sub})}$ and 
$\integralFinalReally_4^{(\text{sub})}$ are related by the Galois symmetry \eqref{eq: Galois action},
\begin{equation}
	(A^{(\text{diag})}_x)_{i2}=\rho_{-5,2}\left[(A^{(\text{diag})}_x)_{i4}\right],\quad 6\le i \le 10,
\end{equation}
consistent with
\begin{equation}
	\integralFinalReally_2^{(\text{sub})}=\rho_{-5,2}\Bigl[\integralFinalReally_4^{(\text{sub})}\Bigr]\,.
\end{equation}
To obtain an $\eps$-factorized differential equation from $A^{(\text{diag})}_x$, we have to add linear combinations of subsector integrals $\integralFinalReally^{(\text{sub})}$ to the topsector integrals $\integralFinal^{(\sectord)}$.
We make an ansatz for such a transformation, and we fix its coefficients by requiring the non-$\eps$-factorizing pieces to vanish order-by-order in $\eps$ without introducing logarithmic terms.
This leads us to a transformation
\begin{equation}	\integralFinalReally=U^{(\text{total})}\integralprec\,,
\end{equation}
where $\integralFinalReally$ satisfies a fully $\eps$-factorized differential equation.
Due to space constraints, we do not show the transformation $U^{(\text{total})}$ here, but rather provide explicit expressions in terms of series expansions in an ancillary file.

\section{Conclusions} 
\label{sec:conclusions}

In this paper, we have presented a method to derive a basis for Feynman integrals in which the differential equation matrix is in $\varepsilon$-factorized form, applicable also in cases where the Picard--Fuchs operator describing the most complicated sector contains an $\varepsilon$-dependent apparent singularity. The method works by allowing for an ansatz for the differential equation matrix which contains non-zero entries on the next-to-next-to-diagonal or beyond, i.e.\ there are entries for which $A_{k,k+2} \neq 0$. As a proof of principle, we obtained an $\varepsilon$-factorized differential equation for the four-loop Feynman integral containing a Calabi--Yau three-variety that contributes to black-hole scattering at fifth order in the post-Minkowskian expansion (5PM) and second self-force order (2SF)~\cite{Frellesvig:2023bbf}.

Having obtained an $\varepsilon$-factorized form for the differential equation, an obvious next step would be to integrate it to obtain an analytical result. Doing so would first require computing the boundary values for all ten master integrals, where a distinction between Feynman propagators and retarded propagators becomes important for an application to classical gravity. Moreover, to compute the contribution of this Calabi--Yau integral to the gravitational-wave observables, we would also need the gravity integrand from the post-Minkowskian expansion. Since such a calculation is beyond the scope of this paper, we leave it for future work.  

Whether or not it would be possible to find a seed integral for the Calabi--Yau sector for which the Picard--Fuchs operator is free of the $\varepsilon$-dependent apparent singularity is an open question. However, our method shows that doing so is not a requirement for obtaining an $\varepsilon$-factorized differential equation. It would nonetheless be interesting to investigate the relation between apparent singularities in the Picard--Fuchs operator and the integrand of the seed integral.

More and more Feynman integrals that lie beyond the polylogarithmic realm are currently being discovered. This includes elliptic and K3-type geometries, as well as Calabi--Yau three-varieties of the type discussed here. Beyond, there are Calabi--Yau varieties of higher dimension, as well as higher genus surfaces. Whether those classes of integrals span the whole space of possible Feynman integrals is an open question to which there is no reason to expect an affirmative answer. Given the huge success of $\varepsilon$-factorized differential equations to obtain analytical results for Feynman integrals, polylogarithmic as well as otherwise, it makes sense to try to extend that success. 
The algorithm presented on the previous pages is a step in that direction, as it does not require previous knowledge of the underlying geometry of the Feynman integral.
It is further agnostic of the space of appropriate forms appearing in the differential equation, as these are solely determined through differential constraints that they satisfy.
Whether our algorithm is sufficient for all Feynman integrals is a question for the future.

\section*{Acknowledgements}

We thank Gustav Jakobsen, Albrecht Klemm, Christoph Nega, Anna-Laura Sattelberger, and Duco van Straten for discussions.

The work of HF, RM and MW was supported by the research grant 00025445 from Villum Fonden. HF has received funding from the European Union's Horizon 2020 research and innovation program under the Marie Sk{\l}odowska-Curie grant agreement No. 847523 `INTERACTIONS'. MW was further supported by the Sapere Aude: DFF-Starting Grant 4251-00029B. This research was in addition supported by the Munich Institute for Astro-, Particle and BioPhysics (MIAPbP) which is funded by the Deutsche Forschungsgemeinschaft (DFG, German Research Foundation) under Germany's Excellence Strategy – EXC-2094 – 390783311.
SP is grateful for the hospitality of Perimeter Institute where part of this work was carried out. 
Research at Perimeter Institute is supported in part by the Government of Canada through the Department of Innovation, Science and Economic Development and by the Province of Ontario through the Ministry of Colleges and Universities. 
This work was supported by a grant from the Simons Foundation (1034867, Dittrich).

\appendix

\section{Picard--Fuchs operator in general dimension}
\label{app:PF_coefficients}

In this appendix, we provide the explicit form for the coefficients of the Picard--Fuchs operator~\eqref{eq:PF_L5} in general dimension $D=4-2\varepsilon$. They are given by
\begin{align}
C_0(x,\varepsilon)= & \, -3 (1 - 2 \varepsilon)^2 (3 + 32 \varepsilon + 4 \varepsilon^2)(1-x^4) \nonumber \\*
& + (71 + 952 \varepsilon + 364 \varepsilon^2 - 20480 \varepsilon^3 + 
 19920 \varepsilon^4 + 1152 \varepsilon^5 + 64 \varepsilon^6) (1-x^4) \, x^2 \nonumber \\*
& -8 (12 + 157 \varepsilon + 461 \varepsilon^2 + 3196 \varepsilon^3 + 
   19872 \varepsilon^4 + 512 \varepsilon^5 + 1584 \varepsilon^6) (1-x^4) \, x^4 \nonumber \\*
& + (71 + 952 \varepsilon + 364 \varepsilon^2 - 20480 \varepsilon^3 + 
 19920 \varepsilon^4 + 1152 \varepsilon^5 + 64 \varepsilon^6)(1-x^4)\, x^6 \nonumber \\*
& -3 (1 - 2 \varepsilon)^2 (3 + 32 \varepsilon + 4 \varepsilon^2) (1-x^4) \, x^8, \\[0.1cm]
C_1(x,\varepsilon)= & \, \, 3 (1 - 2 \varepsilon)^2 (3 + 32 \varepsilon + 4 \varepsilon^2) \, x \nonumber \\*
& - (71 + 952 \varepsilon + 364 \varepsilon^2 - 20480 \varepsilon^3 + 
 19920 \varepsilon^4 + 1152 \varepsilon^5 + 64 \varepsilon^6) \, x^3 \nonumber \\*
& + (333 + 4396 \varepsilon + 15728 \varepsilon^2 + 93264 \varepsilon^3 + 
 247920 \varepsilon^4 + 376064 \varepsilon^5 + 33280 \varepsilon^6) \, x^5 \nonumber \\*
& -2 (549 + 7680 \varepsilon + 31684 \varepsilon^2 + 
   166624 \varepsilon^3 + 466736 \varepsilon^4 + 762496 \varepsilon^5 \nonumber \\*
& + 
   2040256 \varepsilon^6)\, x^7 + (1875 + 29620 \varepsilon + 126096 \varepsilon^2 + 319408 \varepsilon^3 + 
 1509520 \varepsilon^4 \nonumber \\*
& + 578816 \varepsilon^5 + 33280 \varepsilon^6) \, x^9 - (1231 + 18728 \varepsilon + 52684 \varepsilon^2 - 131520 \varepsilon^3 \nonumber \\*
& + 72144 \varepsilon^4 + 2176 \varepsilon^5 + 64 \varepsilon^6) \, x^{11} + (3 + 32 \varepsilon + 4 \varepsilon^2) (61 - 116 \varepsilon + 
   52 \varepsilon^2) \, x^{13}, \\[0.1cm]
C_2(x,\varepsilon)= & \, -2 (1 - 2 \varepsilon + 4 \varepsilon^2) (3 + 32 \varepsilon + 4 \varepsilon^2)(1-x^2) \, x^2 \nonumber \\*
& \, - 4 (17 + 233 \varepsilon + 1176 \varepsilon^2 + 6668 \varepsilon^3 - 
   3376 \varepsilon^4 - 128 \varepsilon^5) (1-x^2) \, x^4 \nonumber \\*
& \, + 4 (193 + 2350 \varepsilon + 4756 \varepsilon^2 + 16808 \varepsilon^3 - 
   736 \varepsilon^4 - 25216 \varepsilon^5) (1-x^2) \, x^6 \nonumber \\*
& \, -8 (283 + 4423 \varepsilon + 18874 \varepsilon^2 + 45908 \varepsilon^3 + 
   150824 \varepsilon^4 + 12608 \varepsilon^5) (1-x^2)\, x^8 \nonumber \\*
& \, + 2 (1121 + 18350 \varepsilon + 67848 \varepsilon^2 - 
   61720 \varepsilon^3 + 19280 \varepsilon^4 + 256 \varepsilon^5) (1-x^2) \, x^{10} \nonumber \\*
& \, -4 (3 + 32 \varepsilon + 4 \varepsilon^2) (35 - 37 \varepsilon + 
   8 \varepsilon^2) (1-x^2) \, x^{12}, \\[0.1cm]
C_3(x,\varepsilon)= & \, (1 + 4 \varepsilon^2) (3 + 32 \varepsilon + 4 \varepsilon^2) (1-x^2)^2 \, x^3 \nonumber \\*
& - 8 (21 + 239 \varepsilon + 202 \varepsilon^2 + 356 \varepsilon^3 + 520 \varepsilon^4) (1-x^2)^2 \, x^5 \nonumber \\*
& \, + 2 (355 + 5088 \varepsilon + 18056 \varepsilon^2 + 44288 \varepsilon^3 + 
   98224 \varepsilon^4) (1-x^2)^2 \, x^7 \nonumber \\*
& \, -8 (138 + 2363 \varepsilon + 9814 \varepsilon^2 - 4396 \varepsilon^3 + 520 \varepsilon^4) (1-x^2)^2\, x^9 \nonumber \\*
& \, + (85 - 48 \varepsilon + 4 \varepsilon^2) (3 + 32 \varepsilon + 
   4 \varepsilon^2) (1-x^2)^2 \, x^{11}, \\[0.1cm]
C_4(x,\varepsilon)= & \, (3 + 4 \varepsilon) (3 + 32 \varepsilon + 4 \varepsilon^2) (1-x^2)^3 \, x^4 \nonumber \\*
& - (73 + 
 876 \varepsilon + 1452 \varepsilon^2 + 3152 \varepsilon^3) (1-x^2)^3 \, x^6 \nonumber \\*
& \, + (187 + 3284 \varepsilon + 14468 \varepsilon^2 - 3152 \varepsilon^3) (1-x^2)^3 \, x^8 \nonumber \\*
& - (17 - 4 \varepsilon) (3 + 32 \varepsilon + 4 \varepsilon^2) (1-x^2)^3 \, x^{10}.
\end{align}

\bibliography{References}
\bibliographystyle{JHEP}

\end{document}